\documentclass[twocolumn,amsmath,amssymb,nofootinbib,superscriptaddress]{revtex4}

\usepackage{hyperref}
\usepackage{graphicx}
\usepackage{dcolumn}
\usepackage{bm,color}
\usepackage{amssymb}
\usepackage{amsmath}
\usepackage{comment}
\usepackage{makecell,tabularx,multirow}
\usepackage{mathtools}
\usepackage{bigints}
\usepackage{enumitem}

\newcommand{\udt}[3]{#1^{#2}_{\phantom{#2}#3}}
\newcommand{\udut}[4]{#1^{#2\phantom{#3}#4}_{\phantom{#2}#3\phantom{#4}}}

\newcommand{\dut}[3]{#1_{#2}^{\phantom{#2}#3}}
\newcommand{\dudt}[4]{#1_{#2\phantom{#3}#4}^{\phantom{#2}#3}}

\newcolumntype{C}{>{\centering\arraybackslash}X}
\newcolumntype{x}[1]{>{\centering\arraybackslash\hspace{0pt}}p{#1}}

\renewcommand{\arraystretch}{2.5}

\begin{document}

\title{Cosmological bouncing solutions in \texorpdfstring{$f(T,B)$}{ftb} gravity}

\author{Maria Caruana}
\email{maria.caruana.16@um.edu.mt}
\affiliation{Department of Physics, University of Malta, Malta}

\author{Gabriel Farrugia}
\email{gabriel.farrugia.11@um.edu.mt }
\affiliation{Institute of Space Sciences and Astronomy, University of Malta, Malta}
\affiliation{Department of Physics, University of Malta, Malta}

\author{Jackson Levi Said}
\email{jackson.said@um.edu.mt}
\affiliation{Institute of Space Sciences and Astronomy, University of Malta, Malta}
\affiliation{Department of Physics, University of Malta, Malta}

\date{\today}

\begin{abstract}
Teleparallel Gravity offers the possibility of reformulating gravity in terms of torsion by exchanging the Levi-Civita connection with the Weitzenb\"{o}ck connection which describes torsion rather than curvature. Surprisingly, Teleparallel Gravity can be formulated to be equivalent to general relativity for a appropriate setup. Our interest lies in exploring an extension of this theory in which the Lagrangian takes the form of $f(T,B)$ where $T$ and $B$ are two scalars that characterize the equivalency with general relativity. In this work, we explore the possible of reproducing well-known cosmological bouncing scenarios in the flat Friedmann-Lema\^{i}tre-Robertson-Walker geometry using this approach to gravity. We study the types of gravitational Lagrangians which are capable of reconstructing analytical solutions for symmetric, oscillatory, superbounce, matter bounce, and singular bounce settings. These new cosmologically inspired models may have an effect on gravitational phenomena at other cosmological scales.
\end{abstract}

\maketitle

\section{\label{sec:intro} Introduction}

The possibility of cosmological bouncing solutions has attracted a lot of attention in recent years due to the ability of this approach to avoid the unnaturalness of the Universe initiating from a big bang singularity. In these scenarios, cosmic contraction reduces the effective radius of the cosmos to a minimum size which then produces an expanding Universe \cite{Brandenberger:2016vhg,Nojiri:2017ncd,Khoury:2001bz,Steinhardt:2001st,1978SvAL....4...82S,STAROBINSKY198099}. This may also open up possibilities for potential quantum gravity theories in the early Universe such as in Refs.\cite{Ashtekar:2006rx,Biswas:2005qr,Ashtekar:2006wn,Ashtekar:2006uz}. Moreover, apart from preventing an initial singularity, bouncing cosmologies have shown to be a competitive alternative to the standard inflationary paradigm \cite{Mukhanov:2005sc,Bamba:2015uma}, and in some realisations, such as in the matter bounce scenario \cite{Brandenberger:2009yt}, produce a scale-invariant power spectrum similar to inflationary models \cite{Barragan:2009sq,Brandenberger:2012zb,1979ZhPmR..30..719S}.

In the literature, an increasing number of studies concerning potential viable bouncing cosmologies have been explored. Firstly, the relatively recent idea of ekpyrotic/cyclic \cite{Khoury:2001wf,Buchbinder:2007ad} describes a cosmos that cyclically expands and contracts, and has been analysed in $f(R)$ theories of gravity \cite{Nojiri:2011kd}. Other works on this topic range from areas such as scalar-tensor theories \cite{Bamba:2013fha, Odintsov:2014gea,Boisseau:2015hqa} among others to unimodular theories \cite{Nojiri:2016ygo}. However, these proposals are not without their problems. For instance in Ref.\cite{1981SvAL....7...36S} it is found that a particular scalar-tensor model produces an unstable evolution due to ghosts when perturbations from an isotropic and homogeneous cosmology is considered.

\begin{figure*}[ht]
\centering
  \includegraphics[width=4.3cm]{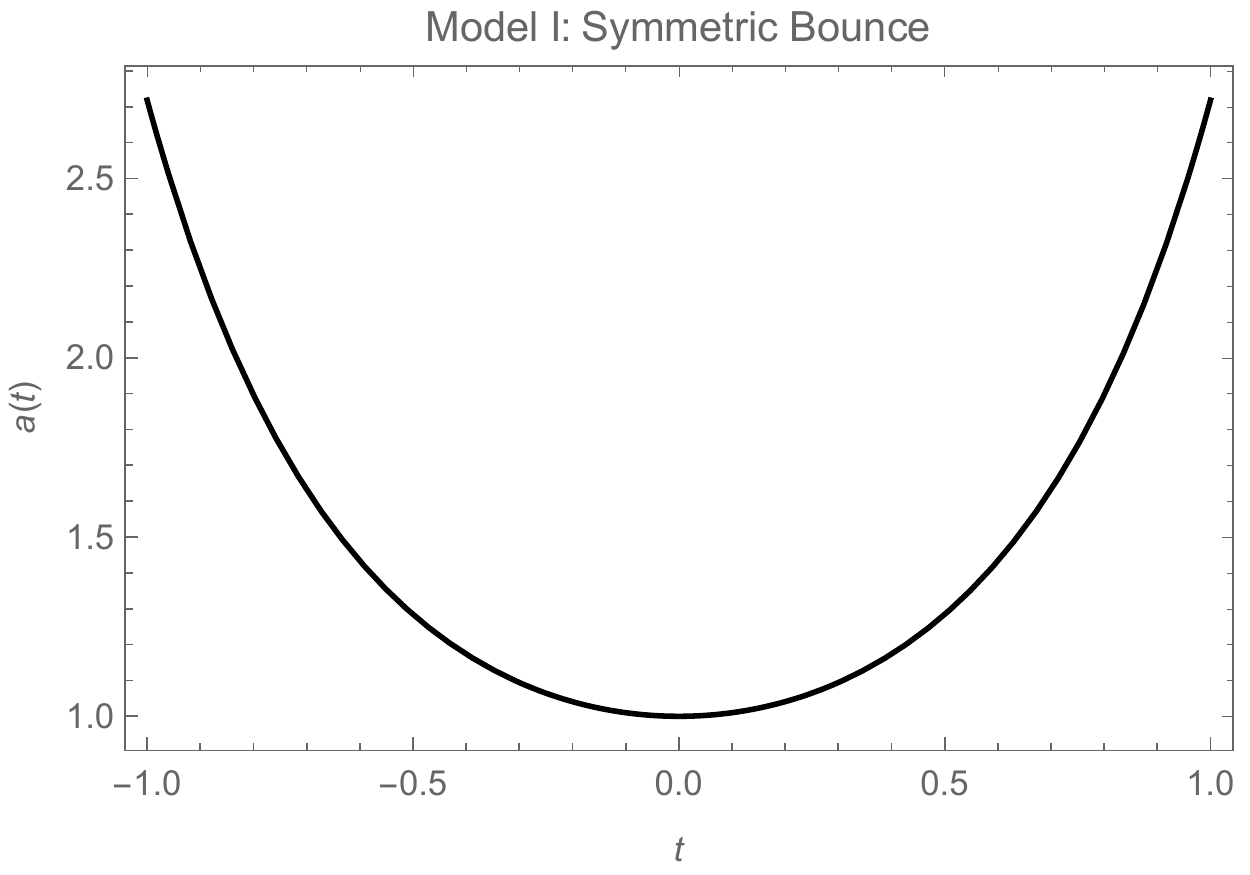}
  \includegraphics[width=4.3cm]{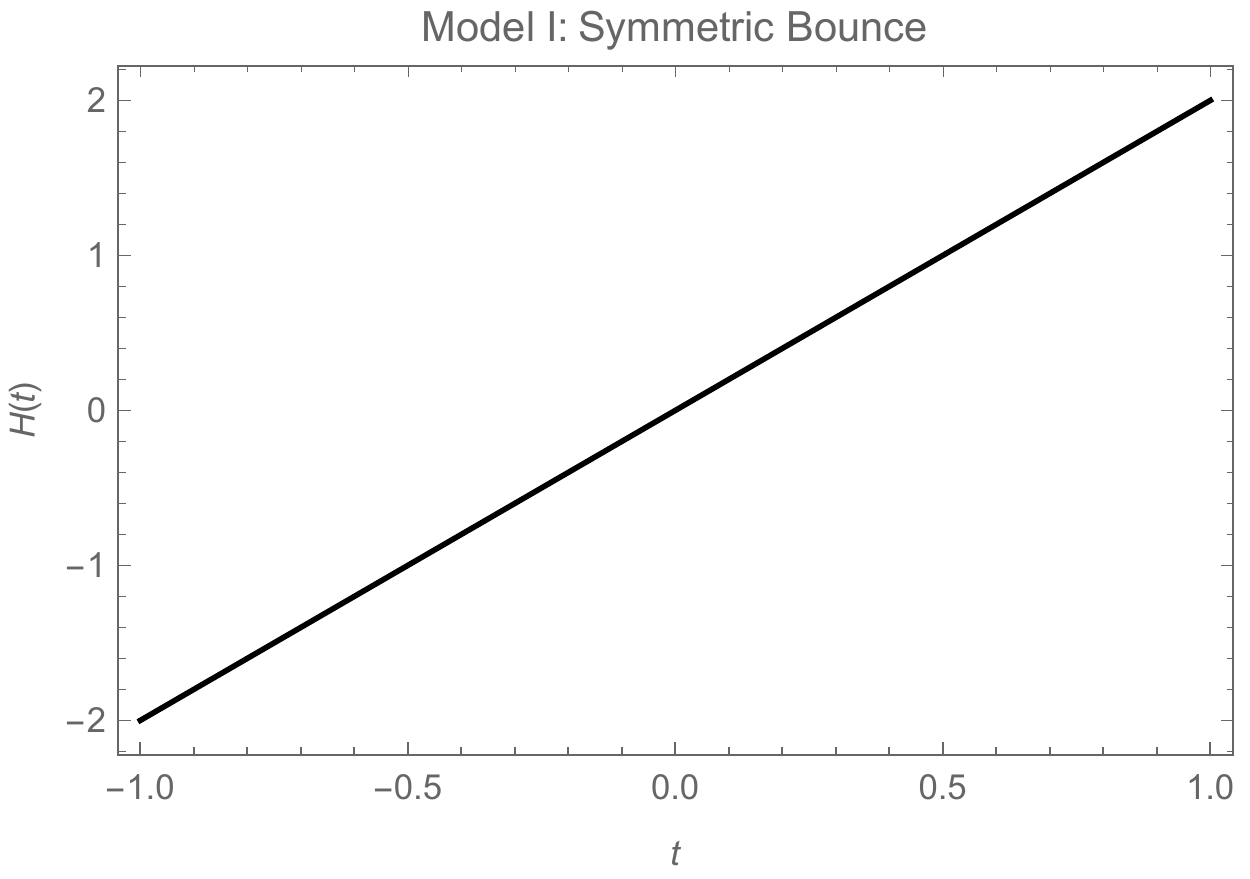}
  \includegraphics[width=4.3cm]{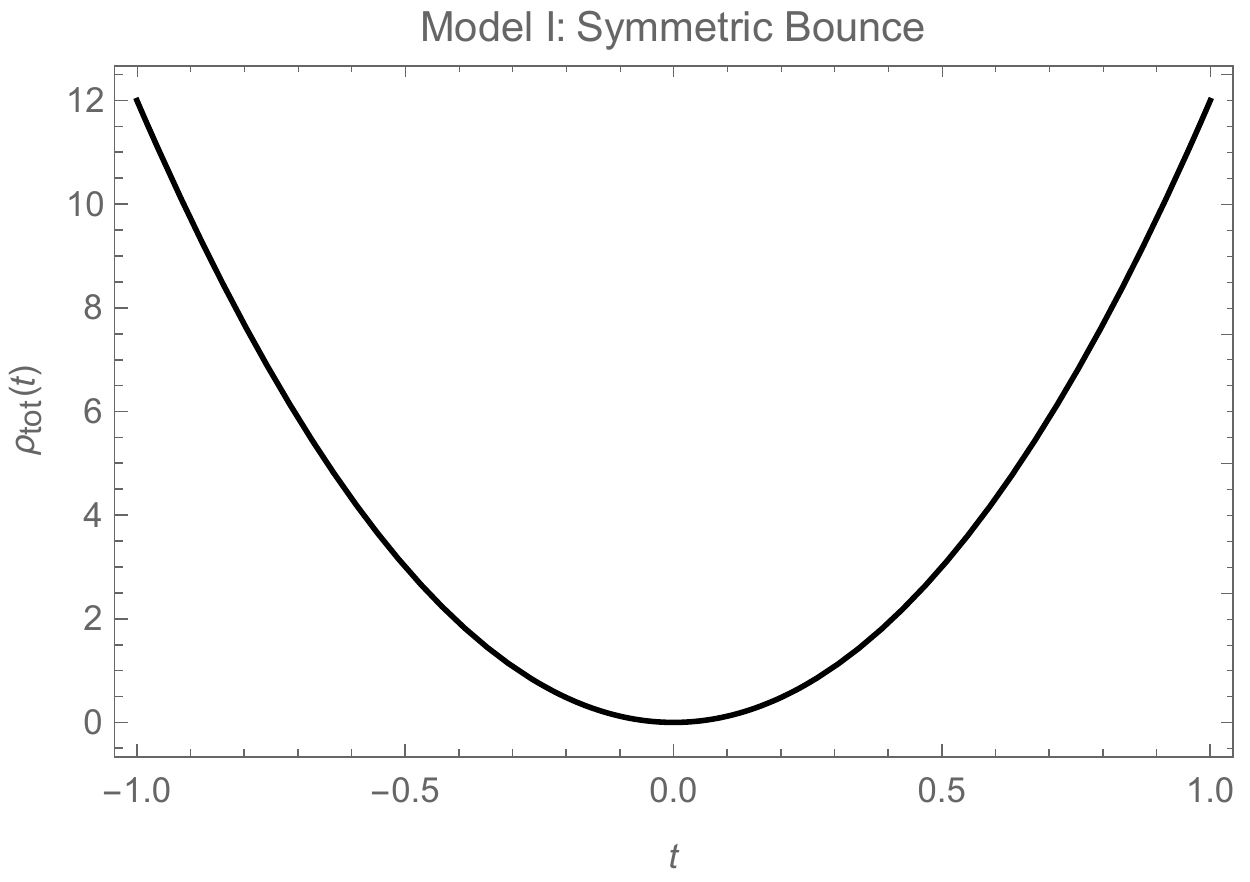}
  \includegraphics[width=4.3cm]{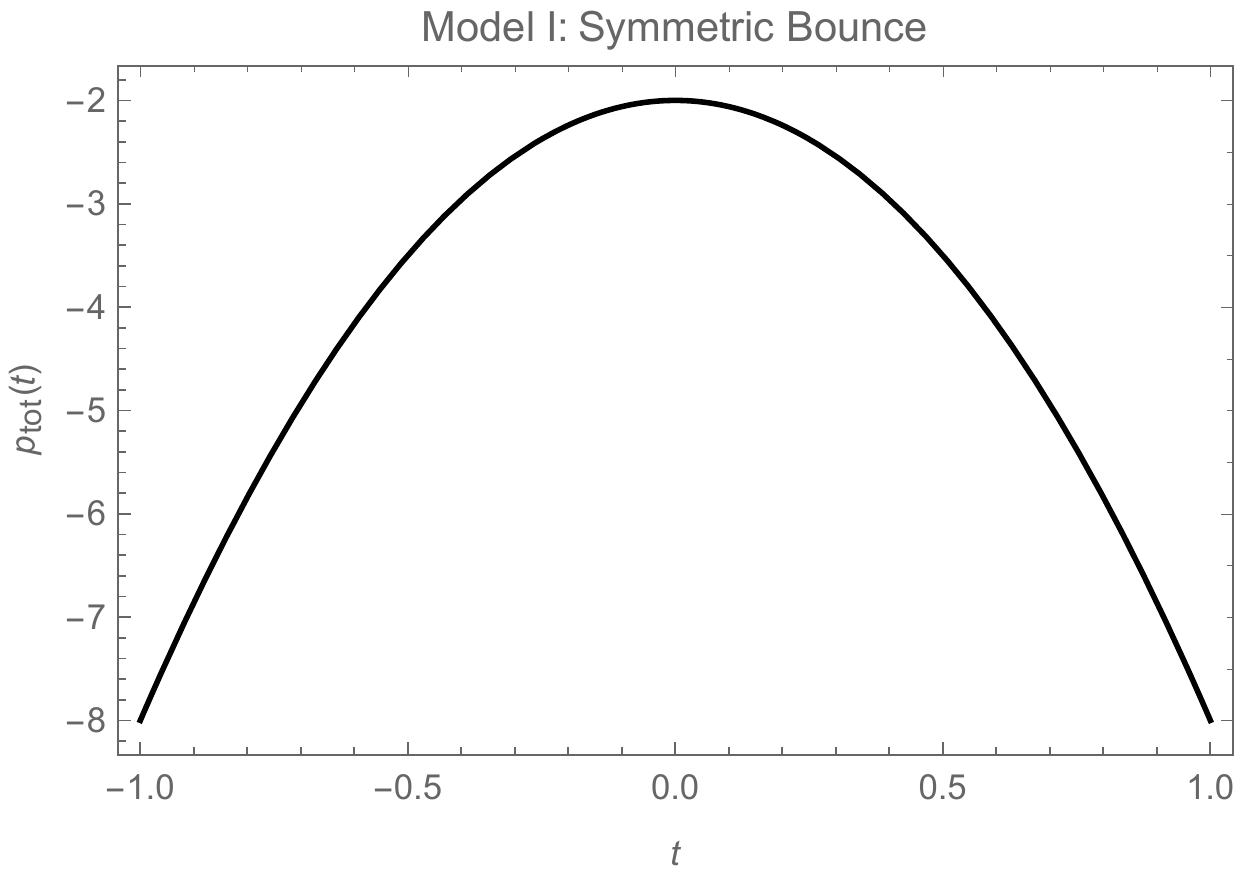}
\hspace{0mm}
  \includegraphics[width=4.3cm]{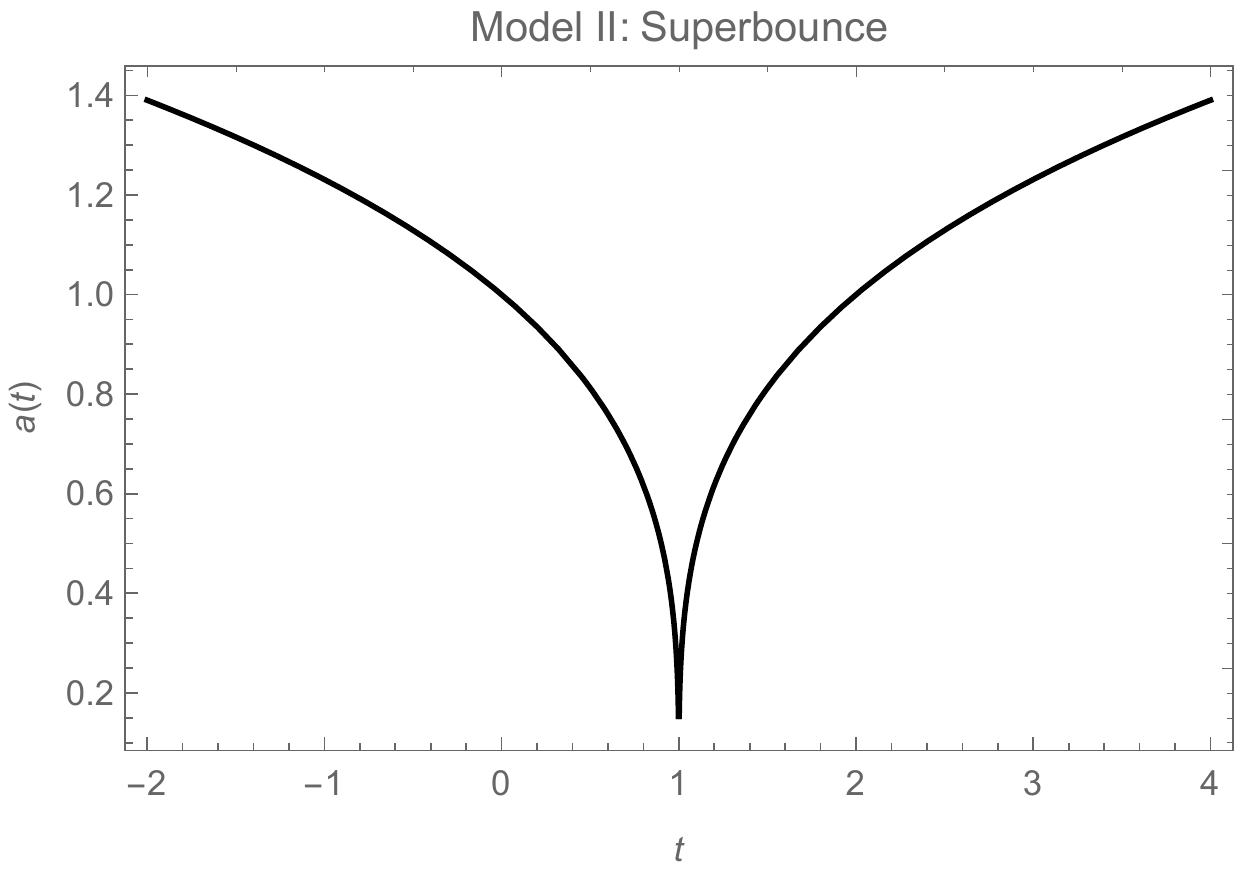}
  \includegraphics[width=4.3cm]{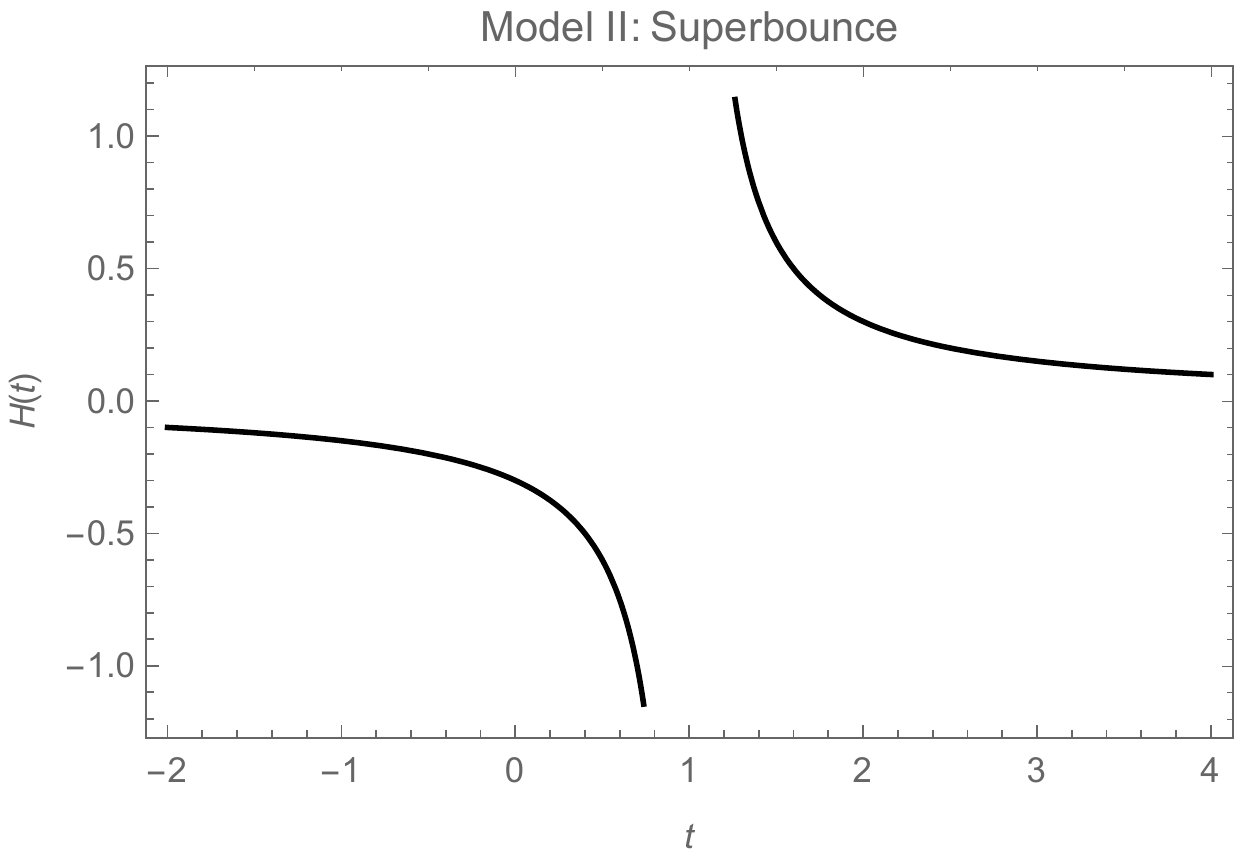}
  \includegraphics[width=4.3cm]{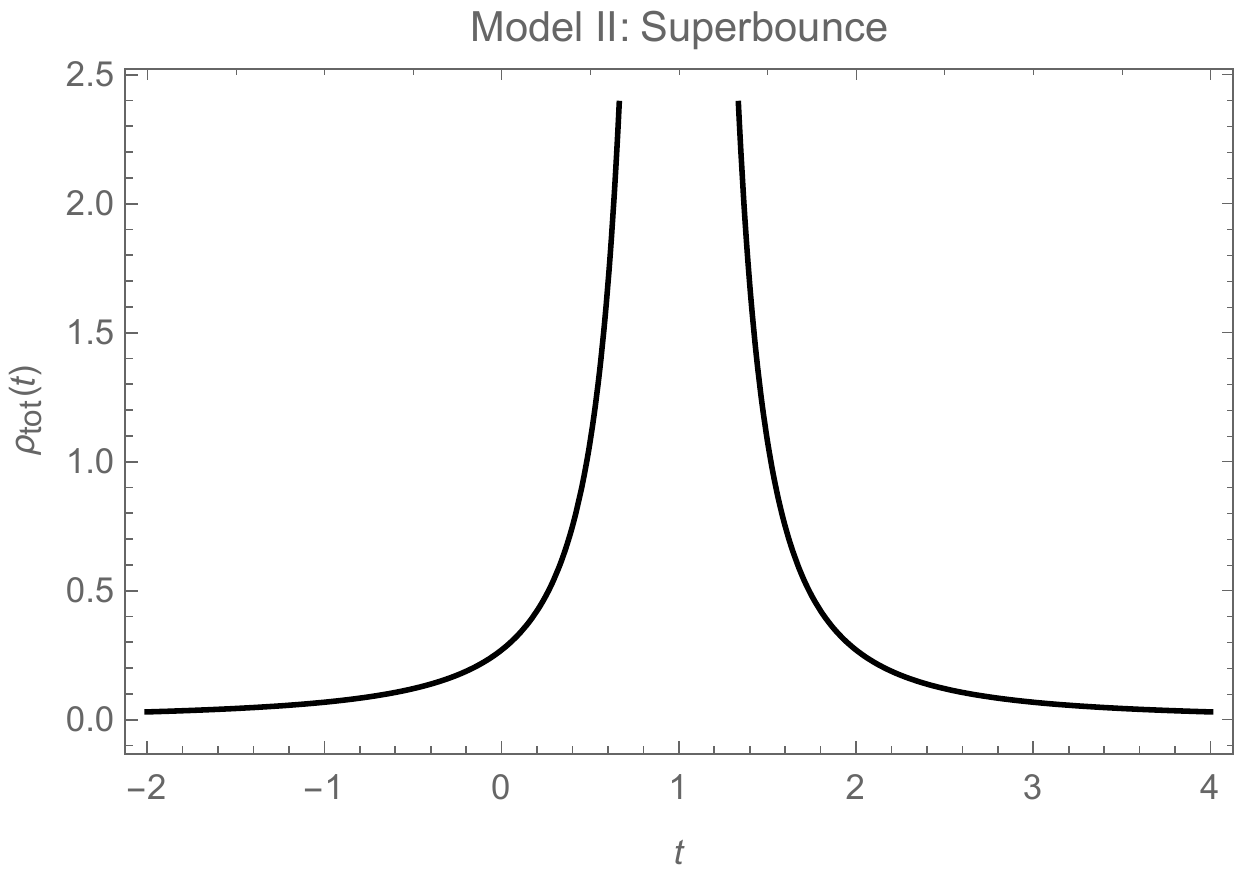}
  \includegraphics[width=4.3cm]{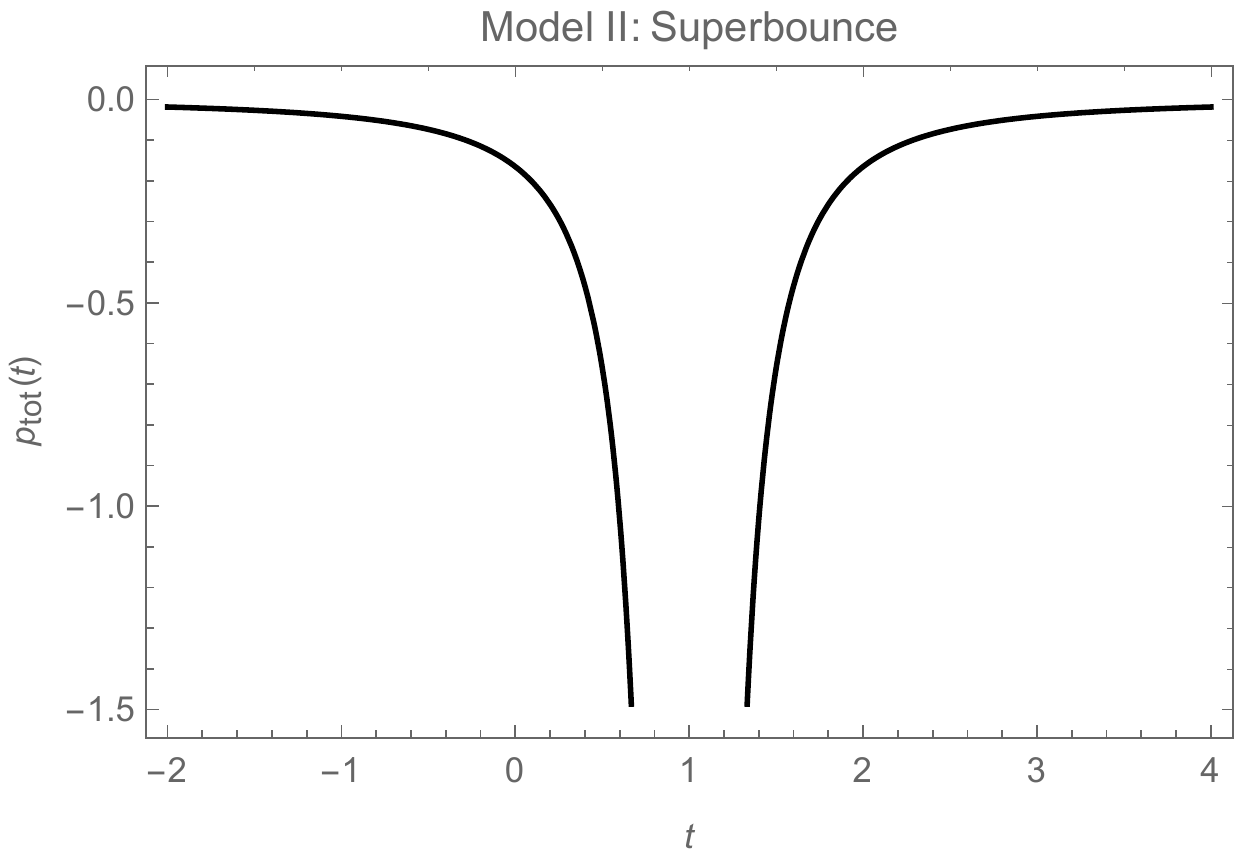}
\hspace{0mm}
  \includegraphics[width=4.3cm]{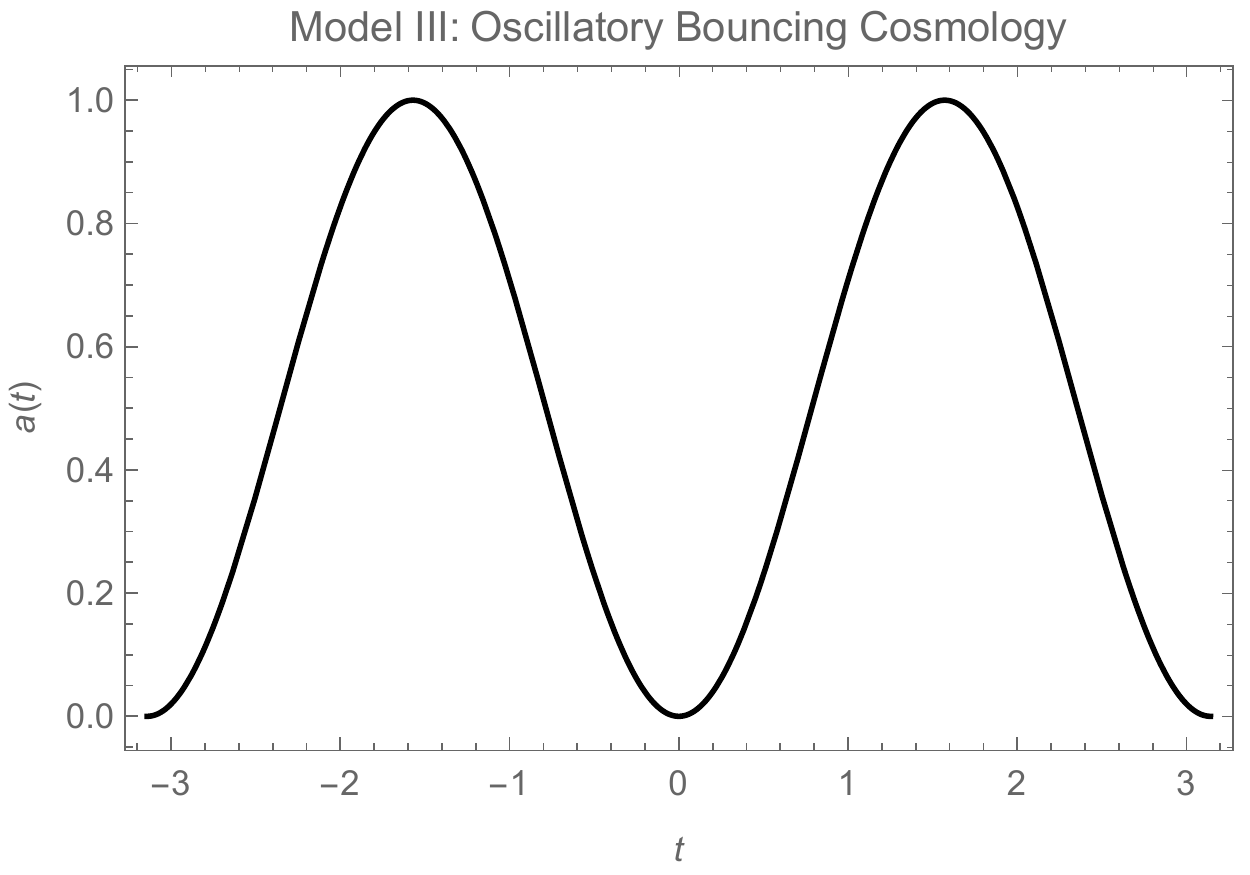}
  \includegraphics[width=4.3cm]{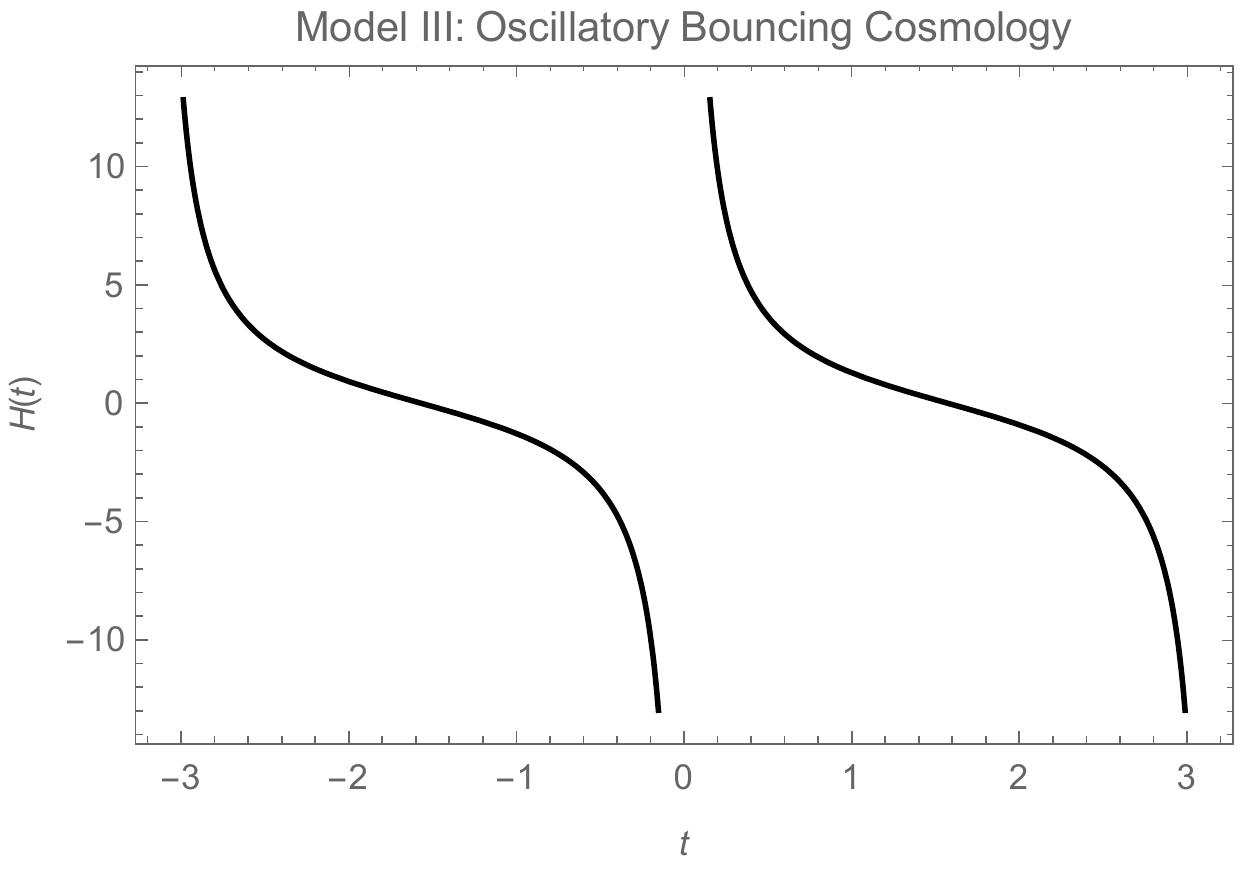}
  \includegraphics[width=4.3cm]{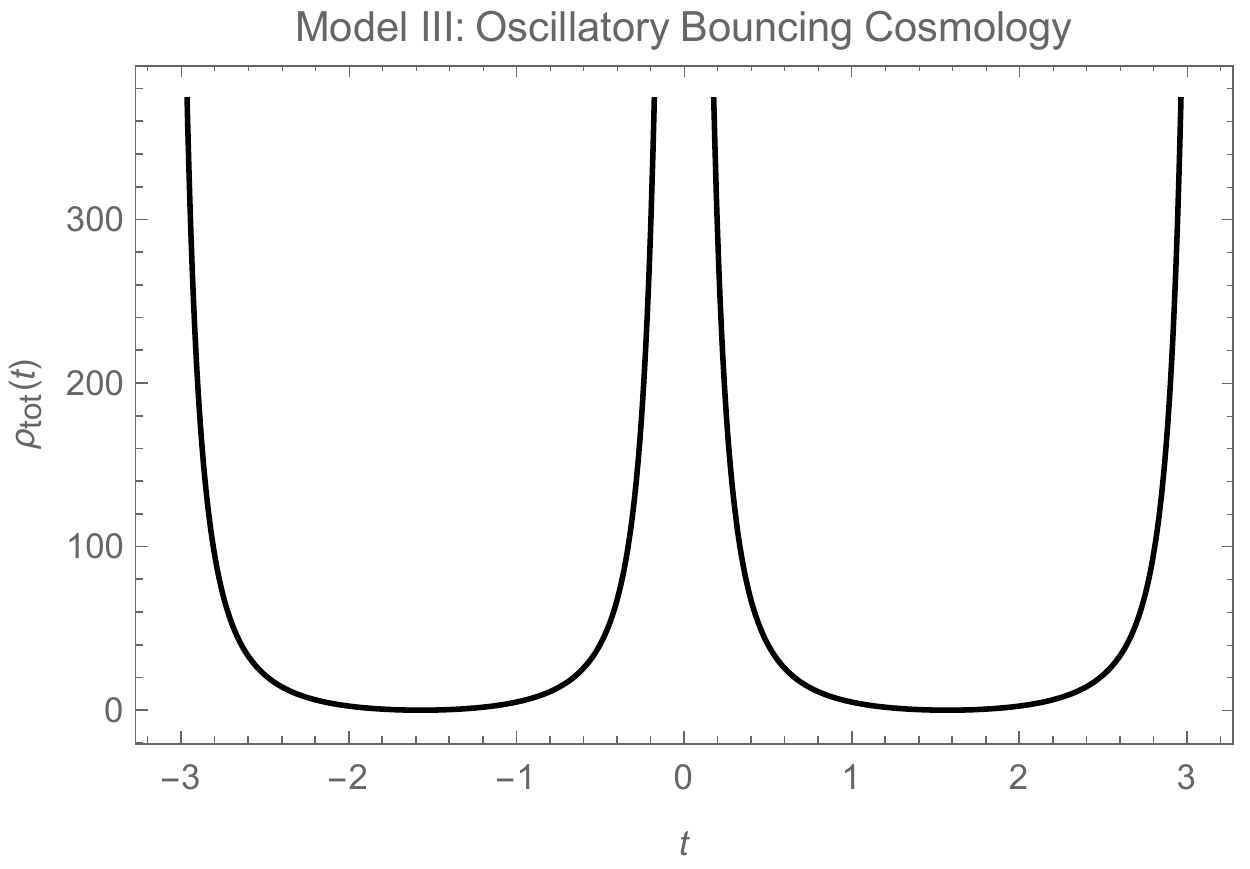}
  \includegraphics[width=4.3cm]{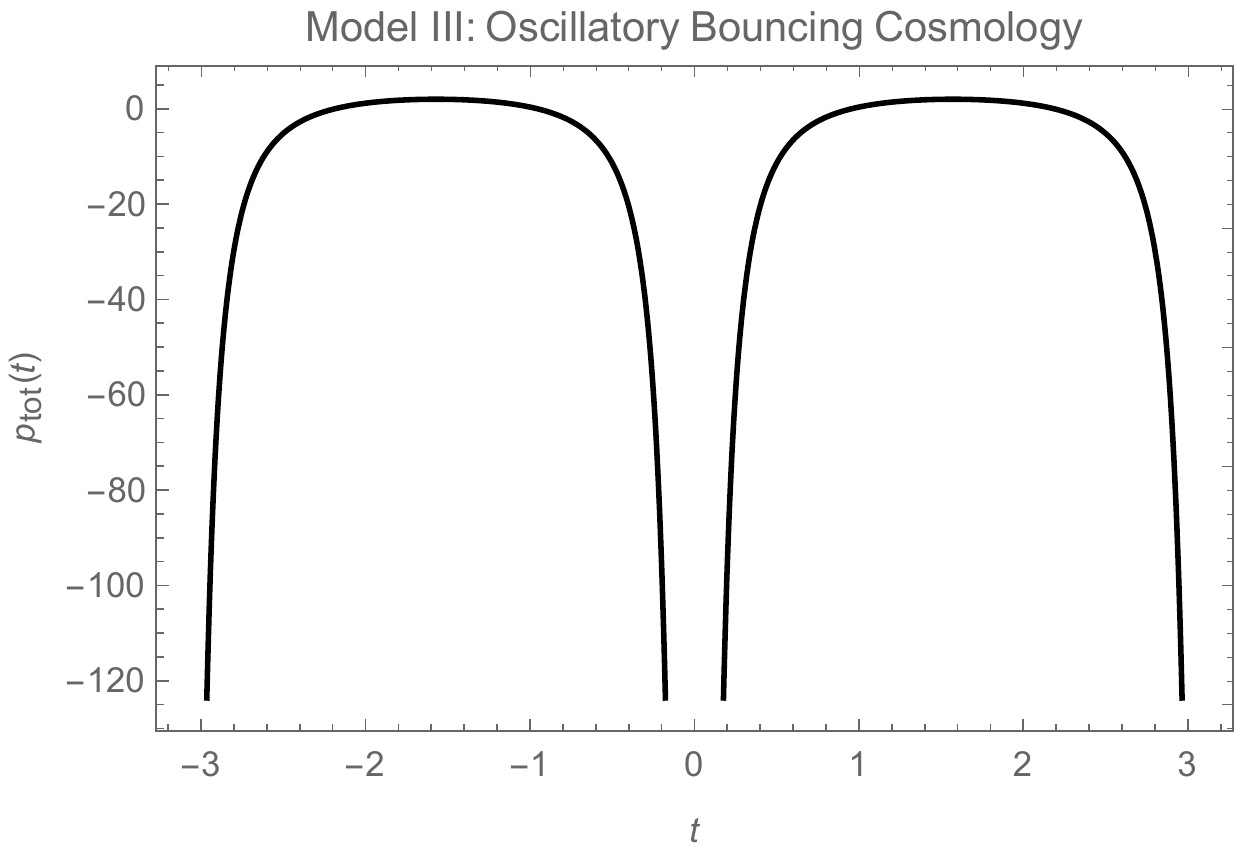}
\hspace{0mm}
  \includegraphics[width=4.3cm]{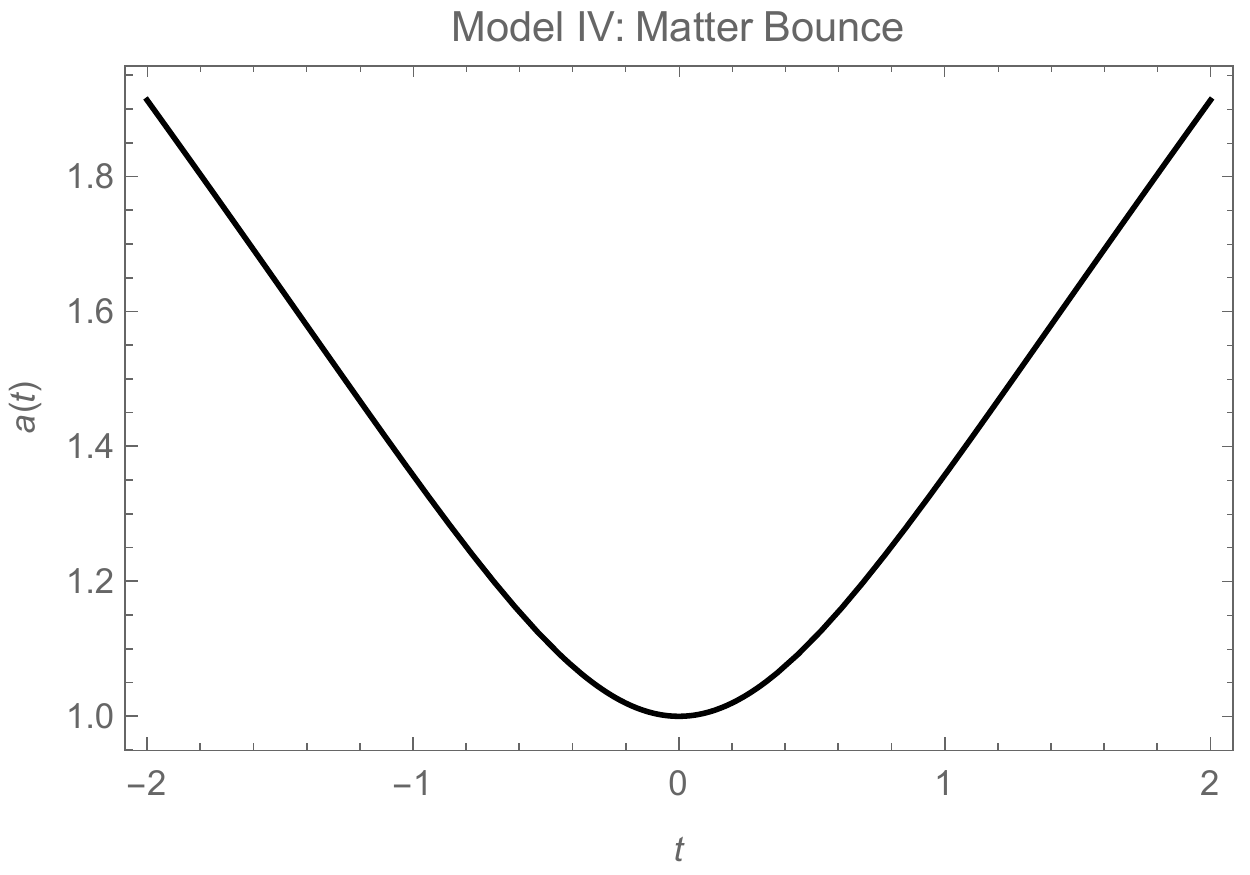}
  \includegraphics[width=4.3cm]{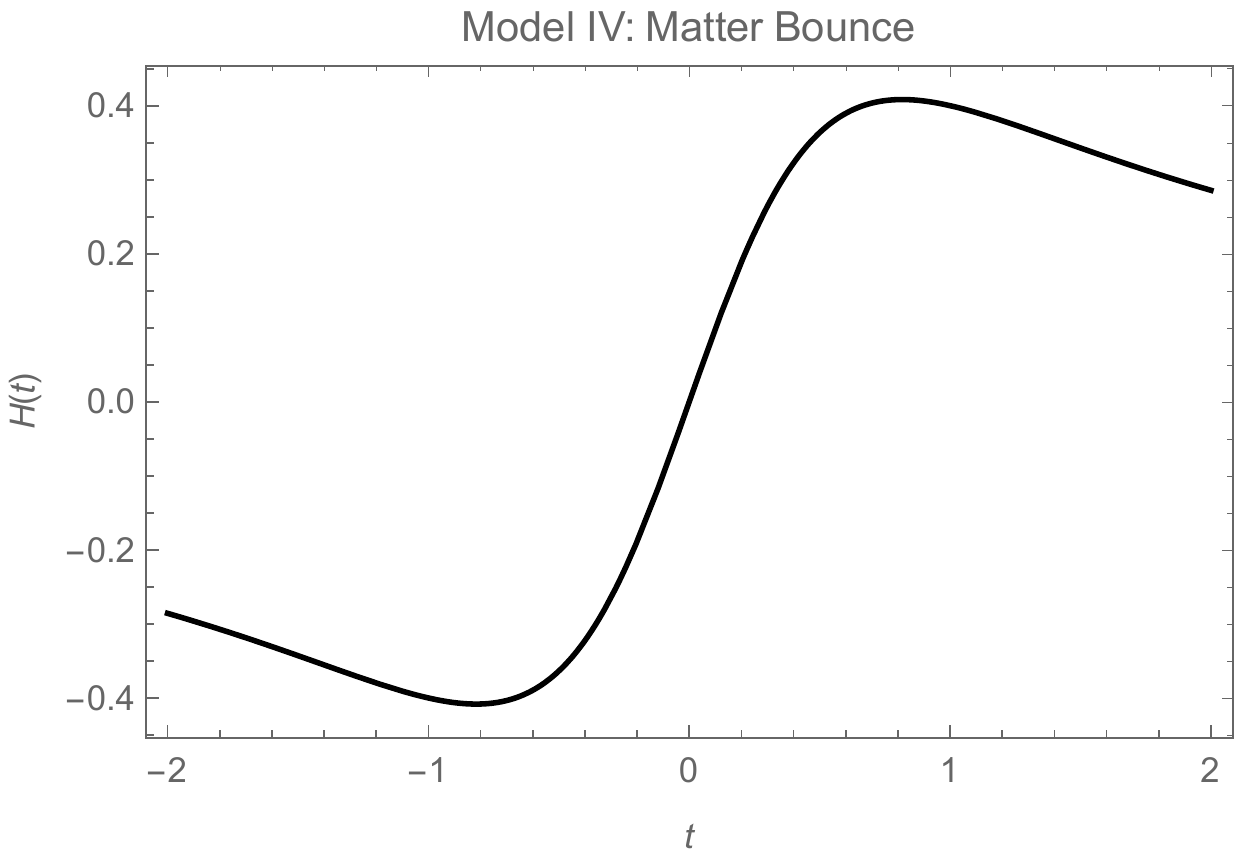}
  \includegraphics[width=4.3cm]{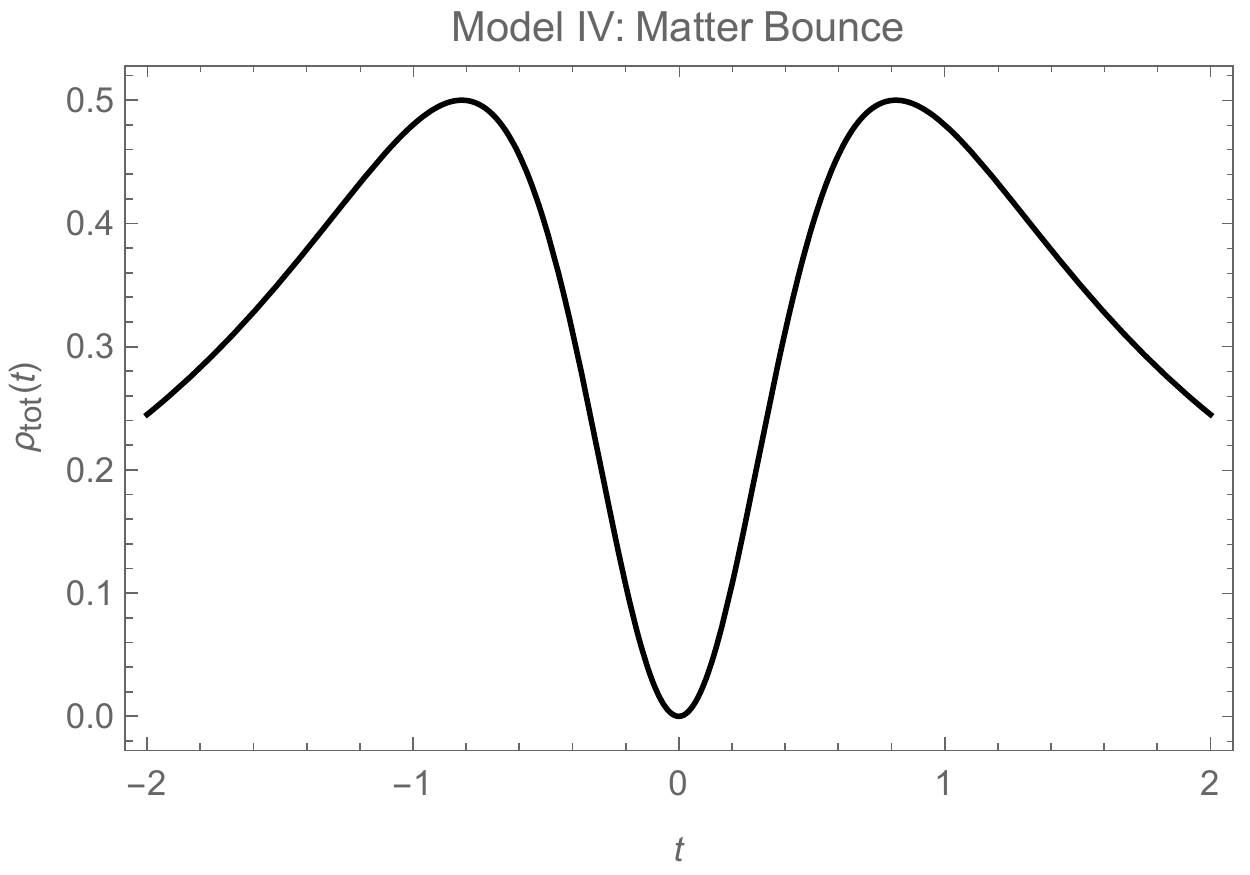}
  \includegraphics[width=4.3cm]{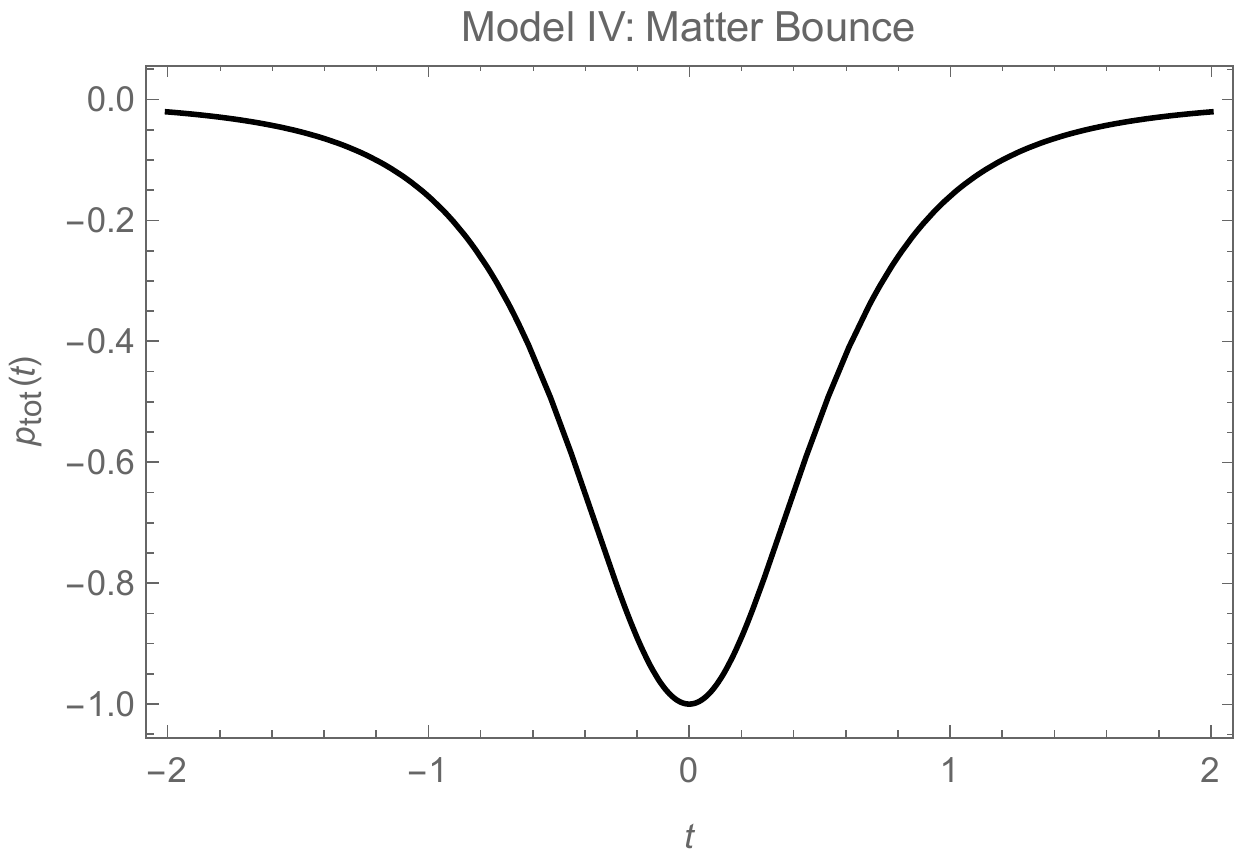}
\hspace{0mm}
  \includegraphics[width=4.3cm]{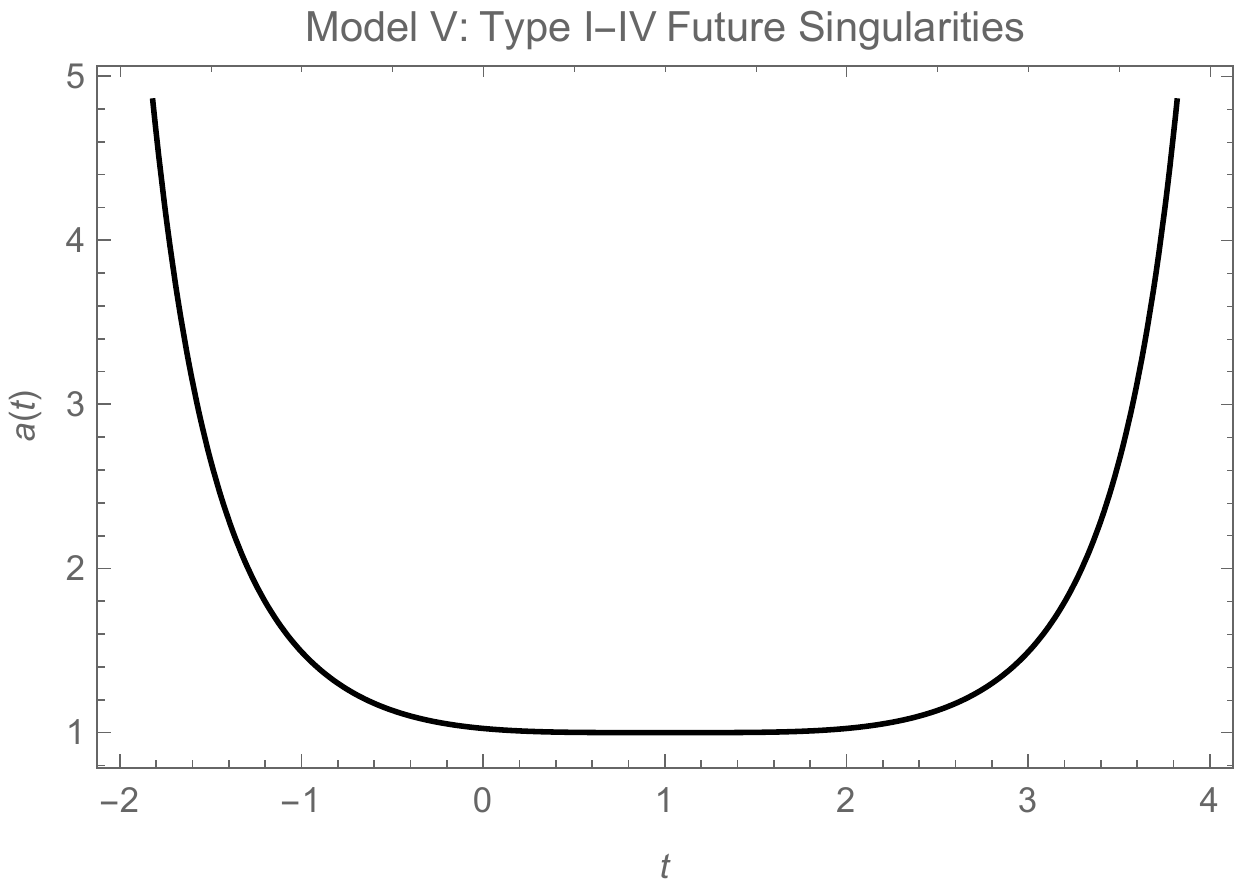}
  \includegraphics[width=4.3cm]{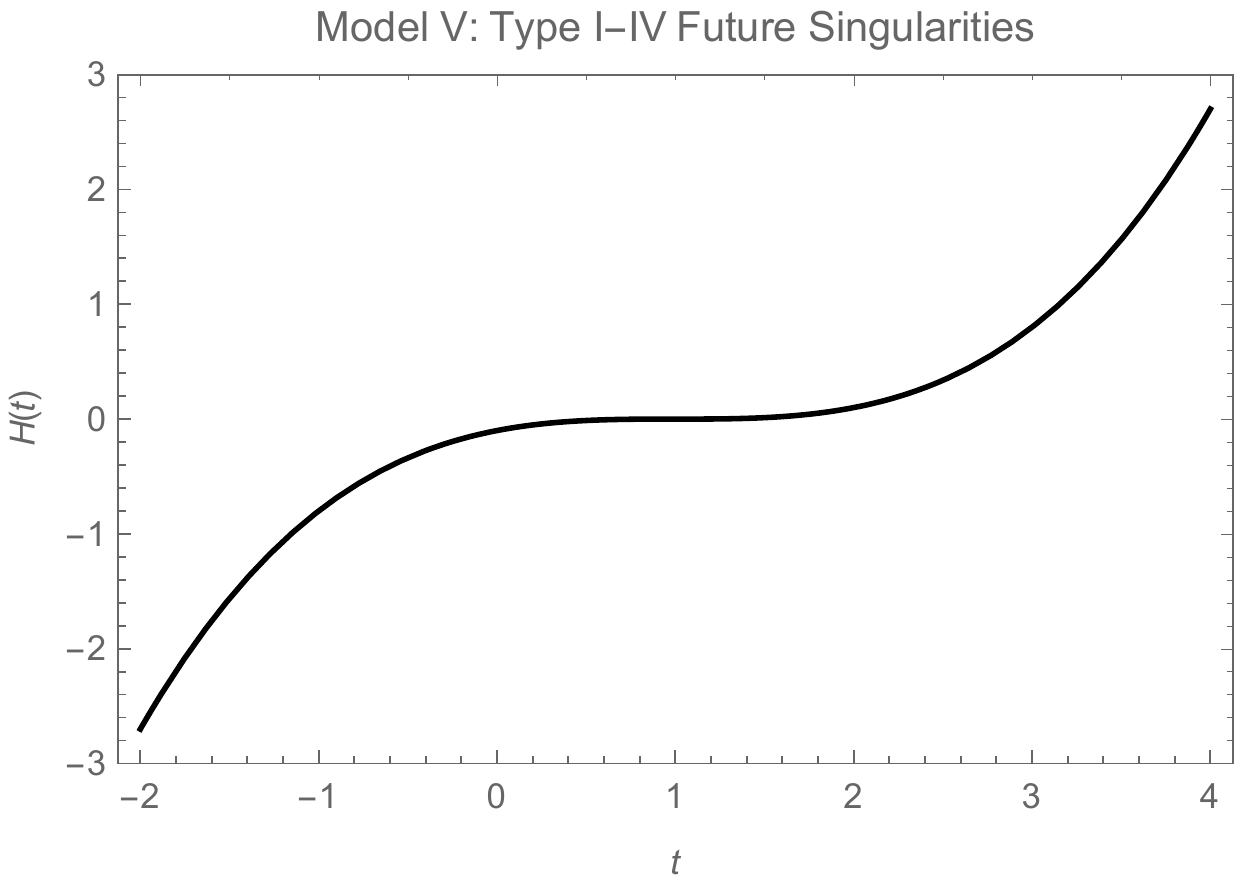}
  \includegraphics[width=4.3cm]{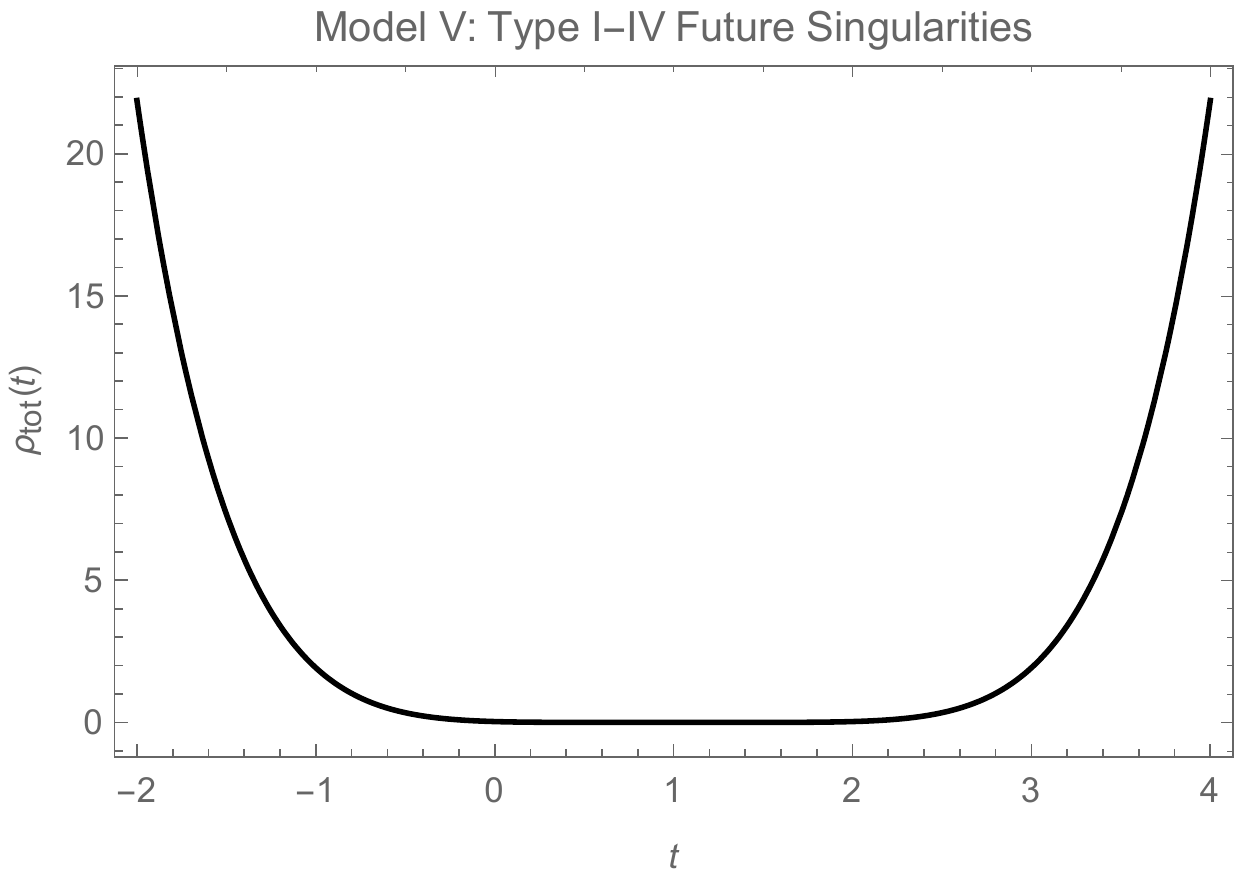}
  \includegraphics[width=4.3cm]{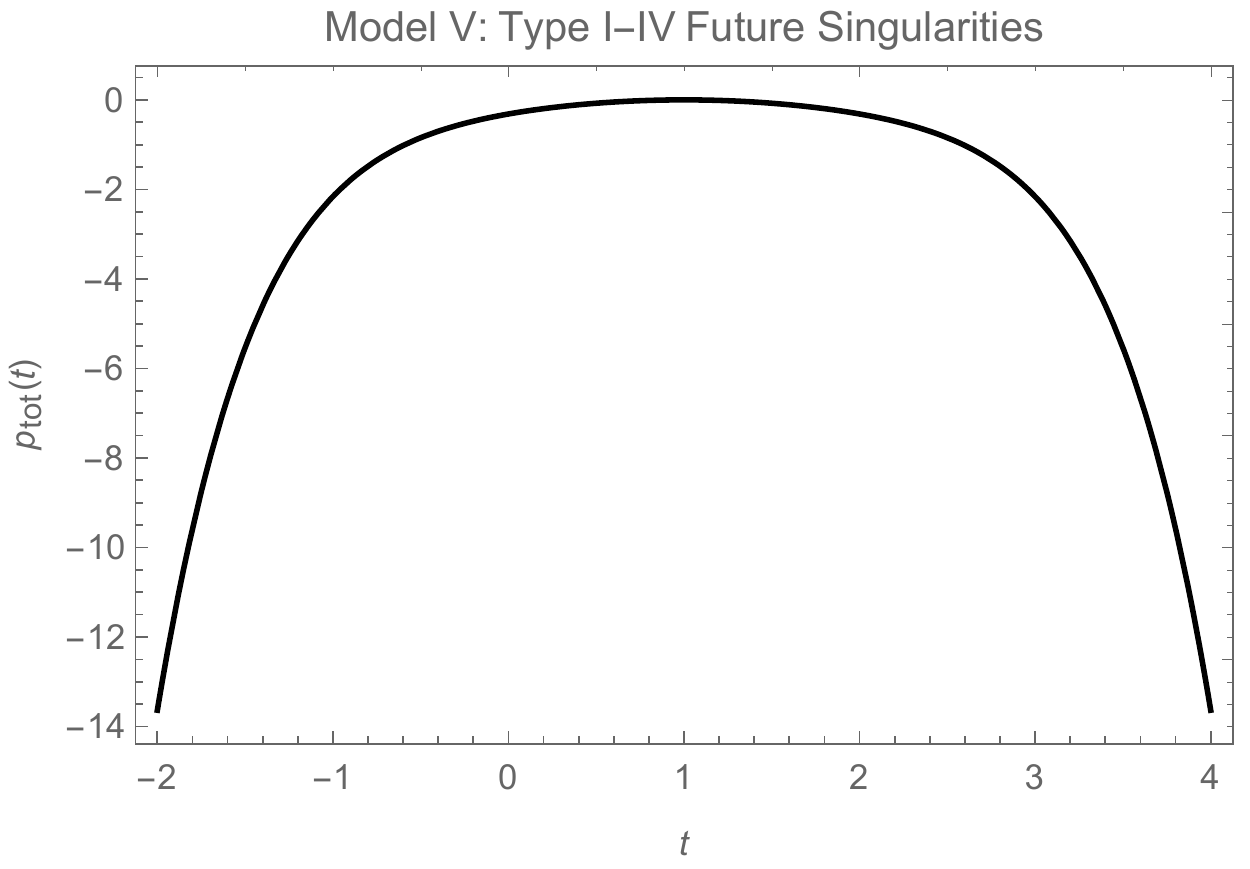}
\caption{For each bouncing scenario analyzed in this work, we plot representative graphs of their scale factor and Hubble parameter, as well as their total density and pressure contributions. The bouncing characteristics are shown as clearly as possible, as are potential singularities (if they occur).  Model V shows a Type IV singularity for which $\alpha = 3$.}
\label{fig:bounce_beha}
\end{figure*}

Superbounce, and ekpyrosis bounce, have also been attracting interest in the literature \cite{Koehn:2013upa,Oikonomou:2014yua,Uzawa:2018sal} with seminal works such as Ref.\cite{Odintsov:2015uca} in which superbounce and the loop quantum cosmology ekyprosis bounce scenarios were investigated for $f(R)$, $f(G)$ and $f(T)$ gravity theories. These effective theories of gravity offer qualitatively similar results indicating a potential universality of this type of bounce scenario. Another interesting proposal for potential bounce cosmologies is that of an oscillatory, or cyclic, bounce \cite{Steinhardt:2001st,Novello:2008ra,Cai:2012va} where a regular periodic bounce occurs at finite temporal intervals, and may offer an new  avenue to resolving cosmological problems in the early universe \cite{Ijjas:2019pyf,Torres:2019kcq}. The differences between these bouncing scenarios can also be viewed through the lens of Fig.\ref{fig:bounce_beha} where the fundamental cosmological parameters are plotted for each bounce model.

The works discussed above and the majority of the literature on bouncing cosmologies is focused on a scenario where general relativity (GR) or its modifications express gravitation. However, another interesting possibility is that of exchanging the fundamental expression of gravity in GR with that of torsion in Teleparallel Gravity (TG). This is achieved by changing the Levi-Civita connection, which is curvature-ful, in GR (and its modifications) with the Weitzenb\"{o}ck connection which is torsion-ful \cite{Weitzenbock1923,RevModPhys.48.393}, but still satisfies the metricity condition. Curvature in GR is expressed not through the metric tensor but through the connection. In this way, TG produces a novel framework in which gravity is realised as a torsional geometric deformation. Thus, we can construct theories of gravity based on the Weitzenb\"{o}ck connection. One such theory is that of the teleparallel equivalent of general relativity (TEGR) which has an associated Lagrangian that is equivalent to GR up to a boundary term \cite{Pereira:2001xf,Maluf:2013gaa}. Therefore, this produces the same dynamical equations as that of GR while being sourced by a different gravitational action.

The boundary term between GR and its TEGR equivalent is the source of many differences in modifications to these theories, which have been studied broadly in the literature \cite{Cai:2015emx,Krssak:2018ywd}. This boundary term arises naturally in GR due to the appearance of second-order derivatives in its Lagrangian. This boundary term \cite{nakahara2003geometry,ortin2004gravity} is the source of the generically fourth-order contributions that arise in extensions to GR \cite{Sotiriou:2008rp,Capozziello:2011et,Clifton:2011jh}. TG features a weakened Lovelock theorem \cite{Lovelock:1971yv,Gonzalez:2015sha,Bahamonde:2019shr} which means that it allows a much wider range of gravitational actions that lead to generically second-order equations of motion. This is a pivotal point for TG since it organically circumvents the appearance of Gauss-Ostrogradski ghosts in many of its manifestations. One interesting use of this is in the formulation of Horndeski theories of gravity within the TG context \cite{Bahamonde:2019shr}.

The TEGR Lagrangian immediately generalizes to produce $f(T)$ theory \cite{Ferraro:2006jd,Ferraro:2008ey,Bengochea:2008gz,Linder:2010py,Chen:2010va} in much the same way that the Einstein-Hilbert action leads directly to $f(R)$ generalizations. In fact, a number of $f(T)$ gravity models have shown promise in both cosmological regimes \cite{Cai:2015emx,Nesseris:2013jea,Farrugia:2016qqe}, as well as for galactic scale physics \cite{Finch:2018gkh} and in solar system tests \cite{Farrugia:2016xcw,Iorio:2012cm,Ruggiero:2015oka,Deng:2018ncg}. However, $f(R)$ gravity is a fourth-order theory and to fully embrace this possibility in the TG context, we must consider $f(T,B)$ gravity theories in which the second-order and fourth-order derivative contribution to the field equations contribute independently to the gravitational action. 

Bouncing scenarios have been investigated in TG, firstly, in terms of its $f(T)$ variant. In this setting, bouncing cosmologies have emerged as a natural consequence in several early universe scenarios, such as the systematic approaches in Refs.\cite{Bamba:2016gbu,ElHanafy:2017xsm,Hohmann:2017jao} show. Others works have also shown the possibility of a matter bounce scenario in $f(T)$ gravity. Beyond $f(T)$ gravity, the literature also includes works on the effect of considering TG as an effective field theory of loop quantum gravity which gives interesting results that are consistent with current observations \cite{Haro:2014wha,Haro:2013bea}. Another aspect of bouncing cosmologies in TG is that of $f(T,T_G)$ where the analog Gauss-Bonnet extension of TG is explored \cite{Kofinas:2014owa,Kofinas:2014aka,Kofinas:2014daa}. This has led to a number of viable models in which bouncing cosmologies can reproduce standard aspects of the early universe \cite{delaCruz-Dombriz:2018nvt,delaCruz-Dombriz:2017lvj}.

Our study explores the possibility of bouncing cosmologies within the $f(T,B)$ gravity framework where we choose to consider the five most studied bouncing cosmology scenarios, namely symmetric bounce, superbounce, oscillatory cosmology, matter bounce, and Type I--IV singularity cases. In fact, some of these models have even been studied for potential cosmological perturbations signatures such as Refs.\cite{Brandenberger:2012zb,Novello:2008ra,Brandenberger:2016vhg,Khoury:2001wf,Cai:2007qw,Odintsov:2014gea}, among others.

In the present work, we investigate the possibility of bouncing solutions within the $f(T,B)$ gravity framework. Our aim is to study several popular manifestations of bouncing cosmologies that appear in the literature (discussed in the following sections). This is achieved by considering a flat Friedmann-Lema\^itre-Robertson-Walker (FLRW) geometry in which the particular forms of the bouncing frameworks emerge through the scale factor. To do this, we first introduce the salient features of TG in \S.~\ref{sec:tg} and discuss relevant features that appear in $f(T,B)$ gravity. In \S.~\ref{exp_scale_fac}, we first investigate the symmetric bounce cosmology, followed by power law models in the context of superbounce scenarios in \S~\ref{power_law_scale_fac}. In \S.~\ref{oscillatory_scale_factor}, a cyclic cosmology described by an oscillating scale factor exhibiting a Big Bang/Big Crunch and a cosmological turnaround bounces are then investigated. Finally, matter bounce and Type I--IV singularity cases are investigated in \S.~\ref{criti_den_scale_fac} and \S.~\ref{exp_2_scale_factor} respectively. The ensuing solutions are discussed in their relevant sections, with a discussion is given in \S.~\ref{sec:conclu}. In this work, we work in units where the speed of light is taken to be unity.

\section{\label{sec:tg} Teleparallel Gravity and its Extension  \texorpdfstring{$f(T,B)$}{ftb} Cosmology}

GR describes gravitation through the Levi-Civita connection, $\mathring{\Gamma}^{\sigma}_{\mu\nu}$ which is curvature-ful and torsion-less while satisfying the metricity condition \cite{misner1973gravitation} (we use overdots throughout to denote quantities calculated with the Levi-Civita connection). TG is centred on the replacement of this connection with a torsion-ful one that has vanishing curvature and satisfies the metricity condition \cite{ortin2004gravity,Cai:2015emx,Krssak:2018ywd}. To achieve this, the Weitzenb\"{o}ck connection, $\Gamma^{\sigma}_{\mu\nu}$, is used to replace the Levi-Civita connection. In GR, the Riemann tensor is used extensively because it gives a measure of curvature on a manifold, and plays an important role in many modified theories of gravity \cite{Clifton:2011jh}. However, by replacing the connection with a curvature-less one implies that the Riemann tensor will always vanish irrespective of the component values of the metric tensor. It is due to this fact that TG requires the bottom-up construction of different tensorial quantities to produce theories of gravity.

The metric tensor, $g_{\mu\nu}$ is the fundamental dynamical object of GR and many of its modifications. However, in TG this is derived from the tetrad, $\udt{e}{a}{\mu}$ which replaces the metric as the acting variable of the theory \cite{Aldrovandi:2013wha}. Here, Latin indices refer to Minkowski space, while Greek indices refer to the general manifold, and the tetrad acts as a soldering agent between the two. In this way, the tetrads (and their inverses $\dut{e}{a}{\mu}$) transform between manifold and Minkowski space indices through
\begin{align}\label{metric_tetrad_eq}
    g_{\mu\nu}=\udt{e}{a}{\mu}\udt{e}{b}{\nu}\eta_{ab}\,,& &\eta_{ab}=\dut{e}{a}{\mu}\dut{e}{b}{\nu}g_{\mu\nu}\,,
\end{align}
which also observe orthogonality conditions
\begin{align}
    \udt{e}{a}{\mu}\dut{e}{b}{\mu}=\delta^b_a\,,& &\udt{e}{a}{\mu}\dut{e}{a}{\nu}=\delta^{\nu}_{\mu}\,,
\end{align}
for consistency. The Weitzenb\"{o}ck connection can then be defined as \cite{Weitzenbock1923}
\begin{equation}
    \Gamma^{\sigma}_{\mu\nu} := \dut{e}{a}{\sigma}\partial_{\mu}\udt{e}{a}{\nu} + \dut{e}{a}{\sigma}\udt{\omega}{a}{b\mu}\udt{e}{b}{\nu}\,,
\end{equation}
where $\udt{\omega}{a}{b\mu}$ represents the spin connection. This constitutes the most general linear affine connection that is both curvature-les and satisfies the metricity condition \cite{Aldrovandi:2013wha}. The spin connection appears explicitly in the Weitzenb\"{o}ck connection to preserve the covariance of the resulting field equations \cite{Krssak:2015oua}. In theories based on the Levi-Civita connection (such as GR), this feature is hidden in the inertial structure of gravity and thus does not play an active role in the ensuing equations of motion \cite{misner1973gravitation,nakahara2003geometry}. The spin connection for TG is flat and incorporates the Local Lorentz Transformation (LLT) invariance of resulting theories. In this way, there will always exist a Lorentz frame where the particular components of the spin connection are allowed to be set to zero.

Considering the full breadth of LLTs (Lorentz boosts and rotations), $\udt{\Lambda}{a}{b}$, the spin connection can be completely represented  as $\udt{\omega}{a}{b\mu}=\udt{\Lambda}{a}{c}\partial_{\mu}\dut{\Lambda}{b}{c}$ \cite{Aldrovandi:2013wha}. Another reason for the spin connection playing an active role in the theory is that for any particular metric, Eq.(\ref{metric_tetrad_eq}) has an infinite number of tetrad solutions, and are each counter-balanced by the spin connection. Thus, it is the tetrad and its associated spin connection that render a covariant TG formulation.

Given a vanishing Riemann tensor for the Weitzenb\"{o}ck connection, we need to replace this with a meaningful measure of torsion. This is achieved through the torsion tensor which takes advantage of the anti-symmetric nature of torsion, defined as \cite{Krssak:2018ywd,Cai:2015emx}
\begin{equation}
    \udt{T}{\sigma}{\mu\nu} := 2\Gamma^{\sigma}_{\left[\mu\nu\right]}\,,
\end{equation}
where square brackets denote the usual anti-symmetric operator. The torsion tensor represents the field strength of gravitation in TG \cite{Aldrovandi:2013wha}, and it transforms covariantly under both diffeomorphisms and LLTs. To formulate interesting theories of gravity, we also necessitate two other quantities. Firstly, consider the contorsion tensor which represents the difference between the Weitzenb\"{o}ck and Levi-Civita connections, i.e.
\begin{equation}
    \udt{K}{\sigma}{\mu\nu} := \Gamma^{\sigma}_{\mu\nu} - \mathring{\Gamma}^{\sigma}_{\mu\nu} = \frac{1}{2}\left(\dudt{T}{\mu}{\sigma}{\nu} + \dudt{T}{\nu}{\sigma}{\mu} - \udt{T}{\sigma}{\mu\nu}\right)\,,
\end{equation}
which plays an important role in relating TG with GR and its modifications. The second central ingredient of TG is the so-called superpotential defined as \cite{Aldrovandi:2013wha}
\begin{equation}
    \dut{S}{a}{\mu\nu} := \frac{1}{2}\left(\udt{K}{\mu\nu}{a} - \dut{e}{a}{\nu}\udt{T}{\alpha\mu}{\alpha} + \dut{e}{a}{\mu}\udt{T}{\alpha\nu}{\alpha}\right)\,,
\end{equation}
which has been shown to have a potential relation to the energy-momentum tensor for gravitation \cite{Aldrovandi:2004db,Koivisto:2019jra}. Contracting the torsion tensor with its superpotential produces the torsion scalar \cite{Cai:2015emx}
\begin{equation}\label{torsion_scalar_def}
    T := \dut{S}{a}{\mu\nu}\udt{T}{a}{\mu\nu}\,,
\end{equation}
which is calculated entirely on the Weitzenb\"{o}ck connection in the same way that the Ricci scalar depends only on the Levi-Civita connection. In the same way, the Ricci scalar as calculated using the Weitzenb\"{o}ck connection will naturally vanish ($R=0$), but using the contorsion tensor this can be related to the regular Levi-Civita calculated Ricci scalar ($\mathring{R}$) through \cite{Hayashi:1979qx,RevModPhys.48.393}
\begin{equation}
    R=\mathring{R} + T - \frac{2}{e}\partial_{\mu}\left(e\udut{T}{\sigma}{\sigma}{\mu}\right) = 0\,.
\end{equation}
This leads to the equivalency between the regular Ricci scalar and the torsion scalar given by
\begin{equation}\label{Ricci_torsion_equiv}
    \mathring{R} = -T + \frac{2}{e}\partial_{\mu}\left(e\udut{T}{\sigma}{\sigma}{\mu}\right) = -T + 2\mathring{\nabla}_{\mu}\left(\udut{T}{\sigma}{\sigma}{\mu}\right)\,,
\end{equation}
where $e=\text{det}\left(\udt{e}{a}{\mu}\right)=\sqrt{-g}$, and $B:=2\mathring{\nabla}_{\mu}\left(\udut{T}{\sigma}{\sigma}{\mu}\right)$ is a boundary term. The appearance of a total divergence term guarantees the equivalence between the dynamical equations that emerge from GR (Ricci scalar Lagrangian) and replacing this with the torsion scalar. Thus, we can define the Teleparallel Gravity equivalent of general relativity (TEGR) as
\begin{equation}
    \mathcal{S}_{\text{TEGR}} = -\frac{1}{2\kappa^2}\int d^4 x\, eT + \int d^4 x\, e\mathcal{L}_m\,,
\end{equation}
where $\kappa^2=8\pi G$ and $\mathcal{L}_m$ is the regular matter Lagrangian. While both Lagrangians lead to the same dynamical equations, their Lagrangians differ by a boundary term that plays an important role in modified versions of GR. In TG, the boundary term embodies the fourth-order derivative contributions to the field equations while in GR, these are contained in the Ricci scalar.

Thus, we can adopt the same reasoning that led to the well-known $f(\mathring{R})$ gravity in the Levi-Civita connection context \cite{Sotiriou:2008rp,Capozziello:2011et}, but in this circumstance, we have two contributing scalars, namely $T$ and $B$. The torsion scalar and boundary term exhibit the second-order and fourth-order derivative contributions respectively. For this reason, we need to generalize to a Lagrangian $f(T,B)$ to suitably incorporate $f(\mathring{R})$ gravity.

Limiting briefly to $f(T)$ gravity, this then produces generally second-order equations of motion unlike its $f(\mathring{R})$ gravity counter-part. This occurs due to a weakening of Lovelock's theorem in TG \cite{Gonzalez:2015sha,Bahamonde:2019shr,Lovelock:1971yv}. Moreover, $f(T)$ gravity shares several other properties with GR such as having the same polarization structure for gravitational waves \cite{Farrugia:2018gyz,Capozziello:2018qcp}, and being Gauss-Ostrogradsky ghost free \cite{ortin2004gravity,Krssak:2018ywd}.

On the other hand, $f(T,B)$ gravity \cite{Bahamonde:2015zma,Capozziello:2018qcp,Bahamonde:2016grb,Paliathanasis:2017flf,Farrugia:2018gyz,Bahamonde:2016cul,Bahamonde:2016cul,Wright:2016ayu} acts as a novel approach to modifying gravity which limits to $f(\mathring{R})$ gravity when the arguments take the specific form $f(T,B)=f(-T+B)=f(\mathring{R})$ gravity. By considering this as a modification to the TEGR Lagrangian, i.e.
\begin{equation}\label{f_T_action}
    \mathcal{S}_{f(T,B)} = \frac{1}{2\kappa^2}\int d^4 x\, e\left(-T+f(T,B)\right) + \int d^4 x\, e\mathcal{L}_m\,,
\end{equation}
we can take the variation to arrive at the following field equations \cite{Bahamonde:2015zma,Farrugia:2018gyz}
\begin{align}
    \dut{e}{a}{\lambda}&\Box f_B - \dut{e}{a}{\sigma}\nabla^{\lambda}\nabla_{\sigma} f_B + \frac{1}{2} B f_B \dut{e}{a}{\lambda}\nonumber\\
    &+ 2\dut{S}{a}{\mu\lambda}\left[\partial_{\mu}f_T+\partial_{\mu} f_B\right] + \frac{2}{e} \left(f_T - 1\right) \partial_{\mu}\left(e\dut{S}{a}{\mu\lambda}\right) \nonumber\\
    &- 2 \left(f_T - 1\right) \udt{T}{\sigma}{\mu a}\dut{S}{\sigma}{\lambda\mu} - \frac{1}{2}\left(-T + f\right)\dut{e}{a}{\lambda} = \kappa^2 \dut{\Theta}{a}{\lambda}\,,
\end{align}
where subscripts denote derivatives, and $\dut{\Theta}{\nu\lambda}=\udt{e}{a}{\nu}\dut{\Theta}{a}{\lambda}$ is the regular energy-momentum tensor for matter. These are derived for a zero spin connection since for a flat FLRW cosmology, this is an allowed value \cite{Bahamonde:2015zma,Capozziello:2018qcp,Bahamonde:2016grb,Farrugia:2018gyz}.

In order to probe the cosmology of $f(T,B)$ gravity, we consider the tetrad choice
\begin{equation}
    \udt{e}{a}{\mu}=\textrm{diag}(1,a(t),a(t),a(t))\,,
\end{equation}
where $a(t)$ is the scale factor, and which reproduces the flat homogeneous isotropic FLRW metric
\begin{equation}
    ds^2=-dt^2+a(t)^2(dx^2+dy^2+dz^2)\,,
\end{equation}
through Eq.(\ref{metric_tetrad_eq}). An interesting feature of this choice of tetrad is that this allows for vanishing spin connection components, i.e. $\udt{\omega}{a}{b\mu}=0$ \cite{Krssak:2015oua,Tamanini:2012hg}. Straightforwardly, we can determine the torsion scalar through Eq.(\ref{torsion_scalar_def}) as
\begin{equation}
    T = 6H^2\,,
\end{equation}
while the boundary term turns out to be \mbox{$B = 6\left(3H^2+\dot{H}\right)$}, which together reproduce the Ricci scalar, i.e. $\mathring{R}=-T+B=6\left(\dot{H} + 2H^2\right)$. From here onwards, overdots refer to derivatives with respect to coordinate time $t$. By evaluating the field equations under these conditions results in the Friedmann equations
\begin{align}
3H^2 &= \kappa^2 \left(\rho_\text{m}+\rho_{\text{eff}}\right)\,,\\
3H^2 + 2\dot{H} &= -\kappa^2\left(p_\text{m}+p_{\text{eff}}\right)\,,
\end{align}
where $\rho_\text{m}$ and $p_\text{m}$ respectively represent the energy density and pressure of matter in the Universe, while $f(T,B)$ enters the governing equations as an effective fluid with energy density and pressure given by
\begin{align}
\kappa^2 \rho_{\text{eff}} &= 3H^2\left(3f_B + 2f_T\right) - 3H\dot{f}_B + 3\dot{H}f_B - \frac{1}{2}f\,, \label{eq:friedmann_mod}\\
\kappa^2 p_{\text{eff}} &= \frac{1}{2}f-\left(3H^2+\dot{H}\right)\left(3f_B + 2f_T\right)-2H\dot{f}_T+\ddot{f}_B\,.
\end{align}
The effective fluid also observes the fluid equation \cite{Bahamonde:2016grb}
\begin{equation}
    \dot{\rho}_{\text{eff}}+3H\left(\rho_{\text{eff}}+p_{\text{eff}}\right) = 0\,,
\end{equation}
and can be used to define an effective equation of state (EoS)
\begin{align}\label{EoS_func}
\omega_{\text{eff}} &= \frac{p_{\text{eff}}}{\rho_{\text{eff}}}\\ 
&= -1+\frac{\ddot{f}_B-3H\dot{f}_B-2\dot{H}f_T-2H\dot{f}_T}{3H^2\left(3f_B+2f_T\right)-3H\dot{f}_B+3\dot{H}f_B-\frac{1}{2}f}\,. 
\end{align}
Notice that $\omega_{\text{eff}}=-1$ is recovered for $\Lambda$CDM where $f(T,B)$ takes on the TEGR limit. Furthermore, since we are considering modified gravity as an alternative description to dark energy, the matter fluid EoS is assumed to satisfy the condition $\omega \geq 0$, which includes known fluids such as dust and radiation. 

In the following section, we will use the Friedmann equations to determine the arbitrary Lagrangian function for different settings of scale factor emanating from bouncing cosmology scenarios. We also note that places where the gravitational Lagrangian exhibits $\sqrt{T}$ or linear $B$ contributions are removed since they act as a total divergence term \cite{delaCruz-Dombriz:2017lvj,delaCruz-Dombriz:2018nvt,Bahamonde:2016cul}.

\section{Reconstruction of Bouncing Cosmologies}

In what follows, we consider the reconstruction procedure for various bouncing cosmologies, namely 
\begin{enumerate}[label = \Roman*.]
    \item Symmetric bounce;
    \item Superbounce;
    \item Oscillatory cosmology;
    \item Matter bounce; and,
    \item Type I--IV and Little Rip cosmology.
\end{enumerate} 
This approach allows for the possibility to solve for the gravitational Lagrangian based on a desired cosmology, which is either set through an analytical form of $a(t)$ or $H(t)$, or through cosmological observations (as, for instance, carried out in $f(T)$ gravity Ref.\cite{Capozziello:2017uam}). Either approach, however, has its limitations.

In most scenarios, the behaviour of $a(t)$ or $H(t)$ would be applicable, or known, only during specific periods. This therefore limits the applicability of the reconstructed Lagrangian as it would only suggest its possible approximate form during certain periods \cite{Capozziello:2005ku}. For a more complete picture, the reconstructed Lagrangian has to match the behaviour over an extended number of cosmological epochs either through a combination of different observations or through reconstruction of unification of different epochs as carried out, for instance, in Refs.\cite{Nojiri:2006gh,Nojiri:2006be}.

The reconstructed solutions obtained in this work are to be treated in a similar fashion, namely that they are a representation of the behaviour of the Lagrangian near the bounce point. Thus, the solutions are not necessarily applicable at late times. Nonetheless, they may provide a clearer picture of the gravitational Lagrangian behaviour at early times which is to be then matched with the overall behaviour of the gravitational Lagrangian throughout the whole universe history. For this reason, as considered in Refs.\cite{Bamba:2013fha,Nojiri:2014zqa,Nojiri:2016ygo} and other related works, asymptotic forms of the solutions near the bounce points shall be considered.

However, solving the resulting partial differential equations arising from the $f(T,B)$ Lagrangian does not generate a general solution for the considered bouncing cosmologies. Thus, particular ansatz forms of the $f(T,B)$ function are instead considered. In particular, the following $f(T,B)$ model ansatz have been considered:
\begin{align*}
(i) &\quad g(T) + h(B),  & (iv) &\quad B g(T), \\
(ii) &\quad T g(B), & (v) &\quad \mu \left(\frac{T}{T_0}\right)^\sigma \left(\frac{B}{B_0}\right)^\gamma, \\
(iii) &\quad T + B g(T), &  & 
\end{align*}
where $\mu, \, \sigma$ and $\gamma$ are constants while $T_0$ and $B_0$ represent the values of the torsion and boundary scalars at times when the scale factor is taken to be unity. 

The first additive separable model encompasses vastly different cosmological models, including TEGR \mbox{$(g = h = 0)$}, $\Lambda$CDM $(g + h = 2\Lambda)$, $f(T)$ gravity $(h = 0)$, TEGR with a modification $(g \neq T)$ allowing for the $g(T)$ and $h(B)$ functions to truly represent the behaviour of the effective fluid component, amongst others. The advantage of such models is the fact that the Friedmann equation yields a decoupled system of ordinary differential equations for the $g(T)$ and $h(B)$ functions making the system simpler to solve. Further details are given in Appendix~\ref{app:additive}.

Models $(ii)$ and $(iii)$ revolve on coupling scenarios which act as modification terms to the TEGR Lagrangian. For the second model ansatz, Frobenius and Green's method were repeatedly used to solve the resulting equations. An interesting feature of model (iii) is that using the action in Eq.(\ref{f_T_action}), it follows that for no choice of potential free parameters does TEGR appear as a subset of this model. In addition to extensions of TEGR, it would also be interesting to develop $f(T,B)$ models that have this property and to test them against observational behaviors and data. A detailed discussion is provided in Appendix~\ref{app:Greens-Method}. 

On the other hand, model $(iv)$, while being similar to model $(iii)$, is fundamentally different as the model cannot recover TEGR and therefore is non-trivial. In this way, the resulting Lagrangian would describe cosmological behaviours without invoking TEGR. Despite this conceptual difference, models $(iii)$ and $(iv)$ are indeed related as they only differ by a particular solution. Once the solution of model $(iii)$ is obtained, model $(iv)$ is given to be
\begin{equation}
    \text{Model $(iv) = $ Model $(iii)$} - T + \frac{B \ln T}{6}.
\end{equation}
This minor difference has important implications when vacuum solutions are considered. 

Lastly, the power-law ansatz model offers a simple Lagrangian which encompasses various models depending on the parameter choices of the free parameters $\mu$, $\beta$ and $\gamma$. Some models include $\Lambda$CDM $(\beta = \gamma = 0)$, TEGR rescaling $(\beta = 1, \, \gamma = 0)$, and $f(T)$ power-law models $(\gamma = 0)$ \cite{Bengochea:2008gz}. These models also appear in the study of Noether symmetry \cite{Bahamonde:2016grb}. A detailed explanation in determining the free parameters according to the considered bouncing cosmology is given in Appendix~\ref{app:power-law}.

Beyond these five ansatz models, it is remarked that any bouncing reconstructed solution derived in any sub-class of $f(T,B)$ gravity, namely $f(T)$, $f(B)$ and $f(R)$ gravity, are naturally solutions to the $f(T,B)$ Friedmann equations. Nonetheless, the given ansatz choices allow for other Lagrangnian solutions, those which are not recovered in any sub-case limit, to appear which may be of crucial importance in other cosmological applications.

Faced with the diverse number of models which could be reconstructed, a further constraint could be imposed in order to be able to distinguish between models which could be deemed as being physically viable. One such constraint is by demanding that the gravitational Lagrangian must be able to recover vacuum solutions such as Minkowski spacetime. Equivalently, this means that in the absence of matter, both $T$ and $B$ scalars are null. From the Friedmann equations Eq.\eqref{eq:friedmann_mod}, this imposes the constraint $f(0,0) = 0$ meaning that no cosmological constant emanates from the Lagrangian.\footnote{Similar considerations appear in other gravitational theories, including $f(R)$ \cite{delaCruzDombriz:2006fj}, $f(T)$ \cite{Ferraro:2011ks} and $f(T,T_G)$ \cite{Tretyakov:2016uvv}.} This condition shall be considered to discuss the viability of the reconstructed solutions.

\begin{table*}[!ht]
\centering
\begin{tabularx}{\textwidth}{|x{2.5cm}|C|C|}
    \hline
    \multicolumn{3}{|c|}{Model I: Symmetric Bounce} \\ \hline 
    
    $f(T,B)$ & Solutions & Asymptotic form close to the bounce \\ \hline
    
    $g(T) + h(B)$ & \makecell{$g(T) = T + T_{0} \Omega_{0} A^{-3(1+\omega)} \left[ e^{-x^2} + \sqrt{\pi} x \,\mathrm{erf}(x) \right]$ \\ $h(B) = C_1 \mathrm{L}\left[-\frac{1}{2}, \frac{3}{2}, \frac{1}{2}\left(-1 + \frac{B}{y}\right) \right]$} & \makecell{$g(T) = T + T_{0} \Omega_{0} A^{-3(1+\omega)} \left(1 + x^2 - \frac{x^4}{6} \right)$ \\ $h(B)= C_{1} \left(\frac{17}{14 \pi } + \frac{11 B}{105 \pi  y} + \frac{B^{2}}{70 \pi y^{2}}\right) $} \\ \hline
    
    $T g(B)$ & $\begin{aligned}g(B) &= C_{2} \left[1+\sqrt{\frac{z\pi }{2y}} e^{\frac{z}{2y}} \, \mathrm{erf}\left(\sqrt{\frac{z}{2y}}\right)\right] + C_{3} \sqrt{z} e^{\frac{z}{2y}} \\ &+ 1 + \int G(z,s) h(s) \; ds \end{aligned}$ & $ \begin{aligned} g(B) &= D_{1} \left(1 + \frac{z}{y}\right) + D_{2} \sqrt{z} \\ &+ 1 -\frac{T_0 \Omega_0 A^{-3(1+\omega)} \ln z}{y} \end{aligned}$ \\ \hline
    
    $T + B g(T)$ & $g(T) = \frac{T_{0} \Omega_{0} A^{-3(1+\omega)}}{6 T} \left[e^{-x^2} + x^2 \, \mathrm{Ei}(-x^2) \right]$ & $g(T) = \frac{T_{0} \Omega_{0} A^{-3(1+\omega)}}{6T}$ \\ \hline
    
    $\mu \left(\frac{T}{T_{0}}\right)^{\sigma} \left(\frac{B}{B_{0}}\right)^{\gamma}$ & \multicolumn{2}{c|}{No analytical solutions exist} \\ \hline
\end{tabularx}
\caption{A summary of the reconstructed Lagrangian solutions together with the associated asymptotic forms close to the bounce in the case of the symmetric bounce cosmology. The parameters $x^2 \coloneqq \frac{3 T (1+\omega)}{2y}$, $y \coloneqq \frac{12 \alpha}{t_{*}^{2}}$, and $z \coloneqq B - y$ have been defined in order to simplify the form of the solutions. Here, $C_{1,2,3}$ are integration constants, $\mathrm{erf}(x)$ is the error function and $L_{a}^b(x) \equiv L[a,b,x]$ is Laguerre's function. The Green's function $G(z,s)$ and $h(s)$ are as defined in Appendix~\ref{app:Greens-Method}.}
\label{table: Table 1}
\end{table*}

\subsection{\label{exp_scale_fac}Model I: Symmetric Bounce}

The first bouncing cosmological model is the symmetric bounce which was originally considered in Ref.\cite{Cai:2012va} to generate a non-singular bouncing cosmology post an ekpyrotic contraction phase. However, this bounce needs to be combined with other cosmological behaviours, otherwise, it suffers from issues with primordial modes not entering the Hubble horizon. \cite{Nojiri:2016ygo,Cai:2014bea,Bamba:2013fha}

This bouncing cosmology is characterised by a scale factor
\begin{equation}
    a(t) = A \exp\left( \alpha \frac{t^2}{{t_*}^2} \right)\,,
\end{equation}
where $t_*$ is some arbitrary time, with $A > 0$ and \mbox{$\alpha > 0$} being positive constants.  The Hubble parameter $H$, torsion scalar $T$ and boundary term $B$ take the simple forms
\begin{align}
        &H = \frac{2 \alpha t}{{t_*}^2}, & &T = \frac{24 \alpha^2 t^2}{t_*^4}, & &B = 3T + \frac{12 \alpha}{t_*^2}.
\end{align}
Evidently, the bounce occurs at $t = 0$ with a preceding contracting phase $(t < 0)$ followed by an expansion phase $(t > 0)$. Moreover, the scale factor can be expressed in terms of $T$ as
\begin{equation}
    a(T) = A \exp \left(\frac{T t_*^2}{24 \alpha} \right).
\end{equation}
For simplicity, by setting $a(t_0) = 1$ for some time $t_0 > 0$, we obtain the expression 
\begin{equation}
    t_0 = \sqrt{- \frac{t_*^2}{\alpha} \ln{A}}\,,
\end{equation}
which implies that the value of $A$ has to be restricted within the region $0 < A < 1$. Consequently, we define the torsion scalar $T$ and the density parameter $\Omega$ at this time $t_0$ as
\begin{align}
        &T(t=t_0) = T_0, & &\Omega (t=t_0) = \Omega_0\,.
\end{align}          
This convention shall be applied for the rest of the models. 

With these definitions, for the $f(T,B)$ ansatz models considered, the solutions are henceforth obtained and are summarised in Table~\ref{table: Table 1}. Furthermore, given the nature of application of bouncing cosmologies is mostly during times close to the bounce point, the obtained Lagrangian solutions can be further approximated by investigating their effective functional forms close to the bounce. For the symmetric bounce cosmology, close to the bounce we have $|T| \ll 1$ and $|z| \coloneqq |B-y| \ll 1$.

Overall, the separable and the $T+Bg(T)$ models recover an exact analytical form while the $Tg(B)$ model only yields analytical results in the absence of matter. Furthermore, the power-law ansatz is unable to describe a symmetric bounce cosmology as discussed in detail in Appendix~\ref{app:power-law}. Despite their complicated forms, the asymptotic limits of the Lagrangian reduce to simple expressions. 

Starting with the additive model, at the lowest order, the model effectively reduces to a TEGR rescaling with a cosmological constant. If higher order contributions are considered, the Lagrangian behaves as an expanded power-series in $T$ and $B$. A similar Lagrangian appears in Ref.\cite{Escamilla-Rivera:2019ulu} where it has been used in the context of the $H_0$ tension. This quadratic limiting order behaviour can be compared with the resulting $f(R)$ asymptotic behaviour obtained in Refs.\cite{Bamba:2013fha,Nojiri:2014zqa} \mbox{$f(R) \propto 144\alpha -72R + \alpha^{-2} R^2$}. As the latter has been derived in the absence of matter sources, this solution is to be compared with the $h(B)$ solution. Through the use of the relations \mbox{$R = -T + B = \frac{2B-y}{3}$}, the $f(R)$ Lagrangian effectively behaves as
\begin{align}
    \mathcal{L}_\text{grav.} &\propto \Lambda_0 + \Lambda_1 B + \Lambda_2 B^2,
\end{align}
for constants $\Lambda_{0,1,2}$ leading to the observed quadratic behaviour obtained here.

Finally, for the $Tg(B)$ model case, the Lagrangian behaves as 
\begin{align}
\mathcal{L}_\text{grav.} &= D_2 \sqrt{B-y} - \frac{T_0 \Omega_0 A^{-3(1+\omega)}}{y} T \ln (B-y) \nonumber \\
&\sim T \ln (T),
\end{align}
while the $Bg(T)$ ansatz leads to an effective power-law behaviour $\mathcal{L}_\text{grav.} \propto \frac{B}{T} \propto 1 + \frac{y}{T} \sim T^{-1}$.

When the vacuum constraint is considered, the resulting conditions are summarised in Table~\ref{table:Model-I-Vacuum}. Overall, only the separable $g(T)+h(B)$ ansatz is able to satisfy this constraint as the remaining models are unable to realise the condition or yield a zero nonphysical gravitational Lagrangian. 

\begin{table}[!h]
\begin{tabularx}{\columnwidth}{|x{2.5cm}|C|}
    \hline
    \multicolumn{2}{|c|}{Model I: Symmetric Bounce}\\ \hline

    $f(T,B)$ & Vacuum Solutions Constraints\\ \hline
    
    $g(T) + h(B)$ & $T_{0} \Omega_{0} A^{-3(1+\omega)} = -\frac{17 C_{1}}{14 \pi}$\\ 
    
    $T g(B)$ & Not possible \\
    
    $B g(T)$ & Not possible \\
    
    $T + Bg(T)$ & $\Omega_{0} = 0$ which implies $\mathcal{L}_\text{grav.} = 0$ and therefore nonphysical \\ 
    
    \hline
    
\end{tabularx}
\caption{A summary of the necessary parameter constraints which need to be satisfied if the reconstructed symmetric bounce cosmological $f(T,B)$ models are to realise the vacuum constraint. Overall, only the additive ansatz is able to satisfy this constraint while also realising the cosmology.}
\label{table:Model-I-Vacuum}
\end{table}

\subsection{\label{power_law_scale_fac}Model II: Superbounce}

Superbounce cosmologies, originally considered in \cite{Koehn:2013upa}, are used to construct a universe which collapses and rebirths through a Big Bang without a singularity \cite{Oikonomou:2014yua}. This type of cosmology is described by a power-law scale factor
\begin{equation}
    a(t) = \left(\frac{t_s - t}{t_0}\right)^{\frac{2}{c^2}}\,,
\end{equation}
where $c > \sqrt{6}$ is a constant, $t_s$ stands for the time at which the bounce occurs, and $t_0 > 0$ is an arbitrary time such that when $t = t_s + t_0$, the scale factor has a unitary value. For this model, the Hubble parameter $H$ turns out to be
\begin{equation}
    H = - \frac{2}{c^2} \left(\frac{1}{t_s - t} \right),
\end{equation}
which identifies the bounce to occur at $t = t_s$. Observe that the superbounce is characterised by a Hubble parameter which changes signature pre- and post-bounce but becomes singular at the bounce point. 

The model can be expressed more simply by a coordinate time shift, $t_{*} \coloneqq t - t_{s}$, leading the bounce to occur at $t_* = 0$. By further defining $\alpha \coloneqq \frac{2}{c^2}$, the expressions for the torsion scalar $T$ and boundary term $B$ are given to be
\begin{align}
    &T = \frac{6 \alpha^2}{t_*^2}, & &B = T \left( \frac{3 \alpha - 1}{\alpha} \right),
\end{align}
while the scale factor is simply expressed in terms of the torsion scalar as
\begin{equation}
    a(T) = \left(\frac{T_0}{T} \right)^{\frac{\alpha}{2}}.
\end{equation}
The resulting solutions for the considered $f(T,B)$ gravitational ansatz together are listed in Table~\ref{table: Table 2}. For this particular cosmology, all ansatz choices generate a simple analytical solution given by a power-law or logarithmic contribution. Thus, the asymptotic form of the Lagrangian close to the bounce remains effectively unchanged. 

Furthermore, for the separable additive ansatz, we recover the $g(T)$ solution obtained in Refs.\cite{Myrzakulov:2010vz,Odintsov:2015uca,Bamba:2016wjm,Bahamonde:2016cul,Darabi:2012zh} and as reported from Noether symmetry \cite{Basilakos:2013rua,Wei:2011aa,Dong:2013rea,Sk:2017ucb,Myrzakulov:2012sp,Atazadeh:2011aa,Sadjadi:2012xa}. On the other hand, the $h(B)$ solution is also reported in Refs.\cite{Bahamonde:2016cul,Zubair:2018wyy}. Finally, the power-law model ansatz solution also appears from Noether symmetry \cite{Bahamonde:2016grb}. 

\begin{table*}[!ht]
\centering
\begin{tabularx}{\textwidth}{|x{2.5cm}|C|}
    \hline
    \multicolumn{2}{|c|}{Model II: Superbounce} \\ \hline 
    
    $f(T,B)$ & Solutions \\ \hline
    
    \multirow{2}{*}{$g(T) + h(B)$} & $\begin{aligned}
    &x \neq \frac{1}{2} & &\hspace{1cm} g(T) = T + \frac{ T_0 \Omega_0}{1-2x} \left(\frac{T}{T_{0}}\right)^{x} \\
    &x = \frac{1}{2} & &\hspace{1cm} g(T) = T - \frac{\Omega_{0} \sqrt{T_{0}\,T}}{2} \ln{T} \end{aligned}$ \\
    & $h(B) = C_{4} B^{\frac{1 - 3\alpha}{2}}$  \\
    \hline
    
    $T g(B)$ & $g(B) = 1 + \frac{2(1 - 3\alpha) \Omega_{0}}{q} \left(\frac{B}{B_0}\right)^{x - 1} + C_{5} B^{\frac{1-3\alpha-z}{4}} + C_{6} B^{\frac{1-3\alpha + z}{4}}$ \\  \hline
    
    $T + B g(T)$ & $\begin{aligned}
    &x \neq 1 & &\hspace{1cm} g(T) = \frac{\Omega_{0}}{6 (1-x)} \left(\frac{T}{T_{0}} \right)^{x-1} \\ 
    &x = 1 & &\hspace{1cm} g(T) = -\frac{\Omega_{0}}{6} \ln{T} \end{aligned}$\\\hline
    
    $\mu \left(\frac{T}{T_{0}}\right)^{\sigma} \left(\frac{B}{B_{0}}\right)^{\gamma}$ & For $\Omega_0 \neq 0$ and $\sigma \neq 0$, $\sigma + \gamma = x = 1$,  $\mu = \frac{T_{0} (1-\Omega_{0})}{\sigma}$  \\ 
    
    \hline
    
\end{tabularx}

\caption{A summary of the reconstructed Lagrangian solutions for the superbounce cosmology. Here, $x \coloneqq \frac{3 \alpha (1+\omega)}{2}$, $z \coloneqq \sqrt{3(\alpha - 3)(3 \alpha - 1)}$, and $q \coloneqq 4 + 2x (2x + 3\alpha - 5)$ have been defined for simplicity, while $C_{4,5,6}$ represent constants of integration.}
\label{table: Table 2}
\end{table*}

When vacuum solutions are considered, it is observed that most models trivially satisfy the constraint with the exception of the $Bg(T)$ and $T+Bg(T)$ models which require a further restriction on the parameter $x$, as shown in Table \ref{table:superbounce-vac}. Nonetheless, this shows that $f(T,B)$ gravity serves as a suitable gravitational model capable of describing a superbounce cosmology while retaining vacuum solutions. 

\begin{table}[!ht]
\centering
\begin{tabularx}{\columnwidth}{|x{2.5cm}|C|}
    \hline
    \multicolumn{2}{|c|}{Model II: Superbounce}\\ \hline

    $f(T,B)$ & Vacuum Solutions Constraints \\ \hline
    
    $g(T) + h(B)$ &  Always satisfied \\ 
    
    $T g(B)$ & Always satisfied \\
    
    $B g(T)$ & $\begin{aligned}
    & 0< x < 1 & & \Omega_{0} = 0 \\
    & x \geq 1 & & \text{Always satisfied}
    \end{aligned}$ \\ [2ex]
    
    $T + Bg(T)$ & {\def\arraystretch{1.25}\tabcolsep=5pt
    \begin{tabular}{@{}lp{4 cm}}
    $0 < x < 1$ & $\Omega_{0} = 0$ which implies $\mathcal{L}_\text{grav.} = 0$ and therefore nonphysical \\ 
    $x \geq 1$ &  Always satisfied
    \end{tabular}} \\ 
    
    $\mu \left(\frac{T}{T_{0}}\right)^{\sigma} \left(\frac{B}{B_{0}}\right)^{\gamma}$ & Always satisfied \\ \hline
    
\end{tabularx}
\caption{The summarised constraints whenever the reconstructed superbounce $f(T,B)$ Lagrangian is able to realise vacuum solutions. Overall, all models are able to generate such solutions while also hosting the superbounce cosmology.}
\label{table:superbounce-vac}
\end{table}


    
    
    
    
    
    

\subsection{\label{oscillatory_scale_factor}Model III: Oscillatory Bouncing Cosmology}

The next model is given by an oscillatory scale factor in the form
\begin{equation}
    a(t) = A \sin^2 \left(\frac{Ct}{t_*}\right),
\end{equation}
where $A$ and $C$ are positive constants, and $t_*$ is some reference time, which for sake of convenience is chosen to be $t_* > 0$. This model represents the behaviour of a cyclic universe \cite{tolman1934relativity,Novello:2008ra}, which treats the universe as a continuous sequence of contractions and expansions \cite{Nojiri:2011kd,Barragan:2009sq,Mukherji:2002ft,Cai:2012ag}. For this particular choice of scale factor, two different bouncing behaviours are encountered. 

The first is a singularity which is experienced throughout each cycle when the scale factor becomes zero while the Hubble parameter becomes singular. This bounce, which occurs when $t = \frac{n\pi t_*}{C}$ for $n \in \mathbb{Z}$, corresponds to a Big Crunch/Big Bang singularity, which could be avoided by constructing a non-zero scale factor or through other mechanisms \cite{Barragan:2009sq,Nojiri:2011kd}. 

The second bounce occurs when the universe reaches its maximal size at $t = \frac{(2n+1)\pi t_*}{2C}$ for $n \in \mathbb{Z}$ leading to a cosmological turnaround. This represents the instance when the universe stops expanding and starts to contract towards the Big Crunch singularity \cite{Cattoen:2005dx}.

\begin{table*}[ht]
\small
\centering
\begin{tabularx}{\textwidth}{ |x{2.5cm}|C| }
    \hline
    \multicolumn{2}{|c|}{Model III: Oscillatory Bouncing Cosmology} \\ \hline 
    
    $f(T,B)$ & \makecell{Solutions}  \\ \hline
    
    $g(T)+h(B)$ & $\begin{aligned} g(T) &= T + T_0 \Omega_0 A^{-3(1+\omega)} \left[\frac{1}{5} x^3 \, _2F_1\left(\frac{5}{2},-3 \omega ;\frac{7}{2};x\right)-x^2 \, _2F_1\left(\frac{3}{2},-3 \omega ;\frac{5}{2};x\right) \right. \\ 
    &\left. +3 x \, _2F_1\left(\frac{1}{2},-3 \omega ;\frac{3}{2};x\right)+\, _2F_1\left(-\frac{1}{2},-3 \omega ;\frac{1}{2};x\right)\right] \\ 
    h(B) &= C_{7} \left[\sqrt{B + y} \left(3 B^{2} + 288 B y + 40 y^{2}\right) - \frac{B}{80 \sqrt{5} y^{\frac{7}{2}}}\arctan{\left(\sqrt{\frac{B + y}{5 y}}\right)}\right] 
    \end{aligned}$ \\ \hline  
    
    $T g(B)$ & $g(z) = C_8 \, _2F_1\left(\frac{5-i \sqrt{15}}{4},\frac{5+i \sqrt{15}}{4};\frac{1}{2};-\frac{z}{2 y}\right)+ C_9 \sqrt{\frac{z}{2 y}} \, _2F_1\left(\frac{7-i \sqrt{15}}{4},\frac{7+i \sqrt{15}}{4};\frac{3}{2};-\frac{z}{2 y}\right) + \bigintsss G(z,s) f(s) \, ds$ \\ \hline
    
    $T + B g(T)$ & For $n \in \mathbb{N}$, $g(T) = -\frac{T_{0} \Omega_{0} A^{-3(1+\omega)}}{12y} \times \begin{cases}
    -(1-x)^{1+3\omega} \left[1 - \frac{(1+3\omega)x^2-(5+9\omega)x}{2 + 9\omega (1+\omega)}\right] - \frac{x}{2 + 9\omega (1+\omega)} + 3 (-1)^{3\omega} (1+\omega) \, \beta\left(\frac{1}{x}; -3\omega, 1+3\omega\right) & \omega \neq \frac{n}{3}, \\
    (n+3) \ln T + \sum\limits_{\substack{k = 0 \\ k \neq 1}}^{n+3} \binom{n+3}{k}\frac{1}{k-1}\left(\frac{T}{2y}\right)^{k-1} & \omega = \frac{n}{3}
    \end{cases}$ \\ \hline
    
    $\mu \left(\frac{T}{T_{0}}\right)^{\sigma} \left(\frac{B}{B_{0}}\right)^{\gamma}$ & No analytical solutions exist \\ \hline
    
\end{tabularx}

\caption{A summary of the Lagrangian solutions for the considered model ansatz in the case of an oscillatory bouncing cosmology described by $a(t) \propto \sin^2\left(\frac{Ct}{t_*}\right)$ for constants $C, \, t_* > 0$. For sake of simplicity, in order to simplify the resulting expressions, the parameters $x \coloneqq - \frac{T}{2y}$, $y \coloneqq \frac{12 C^2}{t_*^2}$ and $z \coloneqq \frac{2}{5}(B+y)$ have been defined. Here $C_{7,8,9}$ are integration constants while $_{2}F_{1}[a,b;c;d]$ represents the hypergeometric function which is undefined for $c = n \in \mathbb{Z}^- \cup \lbrace 0 \rbrace$ and $\beta(x;a,b)$ is the incomplete beta function. For this cosmology, the Green's function $G(z,s)$ which appears in the $Tg(B)$ model as defined in Eq.\eqref{eq:greens-func} has $a(s) = \frac{4s}{5}\left(y+\frac{s}{2}\right)$ while $f(s) = 1 -\frac{ T_0 \Omega_0 A^{-3(1+\omega)}}{s} \left(1 + \frac{s}{2y}\right)^{3(1+\omega)}$.}

\label{table: Table 3}

\end{table*}


For this scale factor, the Hubble parameter is given by
\begin{equation}
    H = \frac{2 C}{t_*} \cot\left(\frac{C t}{t_*}\right),
\end{equation}
from which, the forms of $T$ and $B$ result into
\begin{align}
    &T = \frac{24 C^2}{t_*^2} \cot^2{\left(\frac{C t}{t_*}\right)}, & &B = \frac{5T}{2} - \frac{12 C^{2}}{t_{*}^{2}}.
\end{align}
The definition of the scale factor could therefore be expressed in terms of $T$ as
\begin{equation}
    a(T) = \frac{A}{1 + \frac{T t_*^2}{24 C^2}}.
\end{equation}
Given the nature of the scale factor as a model to describe the whole universe's expansion history, one may consider the current time $t_0>0$ defined through $a(t_0) = 1$ as means to constraint the parameters. This constraint is given by
\begin{equation}
    1 = A \sin^2\left(\frac{C t_0}{t_*}\right),
\end{equation}
which, by definition of the sinusoidal function, leads to the conclusion that $0 < A < 1$, which shall be assumed in what follows. A summary of the obtained Lagrangian solutions is listed in Table~\ref{table: Table 3} while the asymptotic behaviour of the model close to the bounce points (i.e. near the Big Crunch/Big Bang singularity and at the cosmological turnaround) appear in Table~\ref{table:oscillatory-asymptotic}.

For the considered model ansatz choices, only the additive and boundary rescaling models yield an analytical solution. In particular, the additive $g(T)$ solution also appears in Ref.\cite{delaCruz-Dombriz:2018nvt}. Furthermore, it is observed that the power-law ansatz model is unable to describe an oscillatory solution. In the $Tg(B)$ model ansatz, complex arguments appear in the hypergeometric function of the homogeneous solution, which may yield a complex Lagrangian. For the range of values of $-\infty < -\frac{z}{2y} < 0$, it was observed that the hypergeometric function is always real. Nonetheless, in the instance where this is not the case, a simple resolution would be to take $C_{8,9} = 0$.

When asymptotic forms are considered, starting with the behaviour close to the cosmological turnaround (i.e. $H(t) \to 0$), we obtain the following. In the additive case, the $g(T)$ leading order behaviour is a rescaling of TEGR with a cosmological constant. If higher order terms are introduced, a power-law series solution is observed. A similar behaviour is observed in $f(R)$ gravity but for an oscillatory scale factor $a(t) \propto \sin t$ \cite{Carloni:2010ph}. Observe that the higher-order torsional contributions have indices $p \geq 2$, which is expected as such models yield a decelerating cosmology corresponding to the behaviour encountered for a cosmological turnaround \cite{Basilakos:2018arq,Mirza:2017vrk,Zhang:2011qp,Wu:2010xk,Bengochea:2008gz,Nesseris:2013jea,Basilakos:2016xob,Hohmann:2017jao}. Indeed, investigating the effective EoS, $\omega_\text{eff}$, during these times yields diverging, positive values. This is also true for the remaining model ansatz solutions. On the other hand, the $h(B)$ lowest order contribution is of order $\sqrt{B}$. 

Lastly, for the $Tg(B)$ and $T+Bg(T)$ model, the resulting limiting behaviours are similar to the ones obtained in the symmetric bounce cosmology being $\mathcal{L}_\text{grav.} \sim T \ln B$ and $\mathcal{L}_\text{grav.} \propto \frac{B}{T}$ respectively.

When the Big Crunch/Bang singularity is considered, the following is observed. In the additive case, \mbox{$\mathcal{L}_\text{grav.} \sim T^{3(1+\omega)} + B^{\frac{5}{2}}$}. Clearly, there is no TEGR term making the model distinguishable from standard power-law models. In fact, this yields an accelerating cosmology, which differs from the turnaround bounce Lagrangian. Here, the torsion scalar index $3(1+\omega) \geq 3$ is positive and due to the absence of the TEGR contribution, it does not result in a decelerating behaviour but an accelerating one \cite{Wei:2011aa,Atazadeh:2011aa,Basilakos:2013rua}. The latter behaviour is expected for times close to this singularity. Similar power-law behaviours are observed for the matter dependent component of the Lagrangian in the remaining ansatz models.

\begin{table*}[!ht]
\centering
\begin{tabularx}{\textwidth}{|x{2.5cm}|C|C|}
    \hline
    \multicolumn{3}{|c|}{Model III: Oscillatory Bounce} \\
    \hline 
    
    $f(T,B)$ & Asymptotic form close to the bounce as $H(t) \rightarrow 0$ & Asymptotic form close to the bounce as $H(t) \rightarrow \infty$\\ \hline 
    
    $g(T)+h(B)$ &  \makecell{$g(T) = T + T_{0} \Omega_{0} A^{-3(1+\omega)} \left[1 -\frac{3(1+\omega)T}{2y}\right]$ \\ $h(B) = \left(\frac{1}{400 y^3}-245 y^2\right) \sqrt{B+y} \; C_7$} & \makecell{$g(T) = T + \frac{T_{0} \Omega_{0} A^{-3(1+\omega)}}{5 + 6\omega} \left(\frac{T}{2y}\right)^{3(1+\omega)}$ \\ $h(B) = 3 B^{\frac{5}{2}} C_7$} \\ \hline
    
    $Tg(B)$ & $\begin{aligned}
    g(z) &= C_8\left(1-\frac{5 z}{2 y}\right) + C_9\sqrt{\frac{z}{2y}} \\
    &+ 1+ \frac{5T_{0} \Omega_{0} A^{-3(1+\omega)}}{2y} \ln z
    \end{aligned}$ & $g(z) = z^{-\frac{5}{4}}\left[ C_8 \sin \left(\frac{\sqrt{15}}{4} \ln z\right) \right. + \left. C_9 \cos \left(\frac{\sqrt{15}}{4} \ln z \right)\right] + 1-\frac{5\Omega_0 T_0 A^{-3 (1+\omega)}}{2y(18 \omega^2 +39\omega+23)}z^{2+3 \omega}$ \\ \hline
    
    $T + Bg(T)$ & $g(T) = \frac{T_{0} \Omega_{0} A^{-3(1+\omega)}}{6T}$ & $g(T) = - \frac{T_{0} \Omega_{0} A^{-3(1+\omega)}}{12y} \frac{1+3\omega}{2 + 9\omega+9\omega^2} \left(\frac{T}{2y}\right)^{2+3\omega}$ \\ \hline
    
\end{tabularx}
    \caption{A summary of the asymptotic forms of the reconstructed solutions for the oscillatory bouncing cosmology described by $a(t) \propto \sin^2\left(\frac{Ct}{t_*}\right)$ for constants $C, \, t_* > 0$. As this cosmology exhibits two distinct bouncing behaviours, the Big Bang/Crunch singularity and the cosmological turnaround, the respective asmpytotic forms close to each bounce are obtained. Here, \mbox{$z \coloneqq \frac{2}{5}(B+y)$}.}
    \label{table:oscillatory-asymptotic}
\end{table*}



Moving on to the vacuum constraint, as shown in Table~\ref{table:oscillatory-vacuum}, only the additive and $Tg(B)$ models are able to satisfy the constraint while still hosting the oscillatory cosmological behaviour in the presence of matter. This is apparent from the asymptotic behaviour of the models close to the cosmological turnaround as the latter models contain contributions of $T$ and $B$ with positive indices. In the particular $Bg(T)$ case, the model can host vacuum solutions only in the absence of matter fluids.

\begin{table}[!ht]
\small
\centering
\begin{tabularx}{\columnwidth}{|x{2.5cm}|C|}
    \hline
    \multicolumn{2}{|c|}{Model III: Oscillatory Bouncing Cosmology} \\ \hline 
    
    $f(T,B)$ & Vacuum Solutions Constraints\\ \hline 
    
    $g(T) + h(B)$ & Holds for $T_{0} \Omega_{0} A^{-3(1+\omega)} = -40 C_{7} y^{\frac{5}{2}}$\\ 
    
    $T g(B)$ & Asymptotic form indicates that the condition is always satisfied  \\ 
    
    $Bg(T)$ & $\Omega_0 = 0$ which leads to $\mathcal{L}_{\text{grav.}} = -T +  \frac{B\ln{T}}{6}$ \\
    
    $T + Bg(T)$ & $\Omega_{0} = 0$ which implies $\mathcal{L}_\text{grav.} = 0$ and therefore nonphysical \\ \hline
    
\end{tabularx}
\caption{A summary of the conditions necessary for the reconstructed oscillatory cosmology Lagrangians to be able to recover vacuum solutions. Only the additive and $Tg(B)$ models are capable of satisfying this constraint.}
\label{table:oscillatory-vacuum}
\end{table}

\begin{table*}[ht]
\small
\centering
\begin{tabularx}{\textwidth}{ |c|C|x{6cm}|  }
    \hline
    \multicolumn{3}{|c|}{Model IV: Matter Bounce} \\
    \hline 
    
    $f(T,B)$ & Solutions & Asymptotic form close to the bounce  \\
    \hline 
    
    \multirow{2}{*}{$g(T) + h(B)$} & $\begin{aligned}g(T) &= T_{0} \Omega_{0} A^{-3(1+\omega)} \sqrt{1 - \frac{x}{2}} \left(\frac{x}{2}\,_2F_1\left[\frac{1}{2}, \frac{1}{2} - \omega,\frac{3}{2}; \frac{x}{2}\right]\right.\\ & \left.+ \,_2F_1\left[-\frac{1}{2}, -\frac{1}{2}-\omega; \frac{1}{2}; \frac{x}{2}\right]\right) +  T \end{aligned}$ & $\begin{aligned} g(T) &= T_{0} \Omega_{0} A^{-3(1+\omega)} \left[1 + \frac{(1+\omega)}{2} x \right. \\ &\left. - \frac{(1+\omega)(4+\omega)}{24} x^2\right] + T \end{aligned}$ \\ 
    & $h(B) =  \frac{(B - 6\rho_{c})^{\frac{3}{2}}}{9 \sqrt{B} \rho_{c}} C_{10}$ & $h(B) = \frac{(B-6 \rho )^{3/2}}{9 \sqrt{6} \rho ^{3/2}} C_{10}$ \\ \hline
    
    $T g(B)$ & $g(z) = C_{11} g_+(z) + C_{12} g_-(z) + \bigintsss G(z,s) f(s)$ where $g_{\pm}(z) =  z^{\frac{-1\pm\sqrt{7} i}{4}}\, _2F_1\left[\frac{-1\pm\sqrt{7}i}{4}, \frac{3\pm\sqrt{7} i}{4};1 \pm \frac{\sqrt{7}i}{2};z\right]$ & \multicolumn{1}{@{\hskip 8pt}p{5.5cm}|}{The homogeneous solution takes a complicated form. Only the asymptotic form of the particular solution is given: $g_\text{part.}(z) = 1 + \frac{T_0 \Omega_0 A^{-3(1+\omega)}}{4 \rho_c} \left[2 + z \ln{(z-1)}\right]$} \\ \hline
    
    $T + B g(T)$ & $\begin{aligned} 
    & \omega \neq n & & g(T) = \frac{\Omega_0 T_0 A^{-3 (1+\omega)}}{24 \rho_{c}}\left(1-\frac{x}{2}\right)^{\omega} \left(\frac{2}{x}-\frac{1}{\omega }-1\right.\\
    & & &  \left.+ \left(1+\frac{1}{\omega}\right) \left(\frac{x}{x-2}\right)^{\omega} \, _2F_1\left[-\omega ,-\omega ;1-\omega ;\frac{2}{x}\right]\right) \\
    &\omega = 0 & & g(T) = \frac{\Omega_{0} T_{0}}{24 A^3 \rho_{c}} \left[\frac{2}{x} - \ln \left(\frac{2}{x}-1\right)\right] \\
    &\omega \geq 1 & & g(T) = \frac{\Omega_0 T_0 A^{-3 (1+\omega)}}{12} \left[\frac{1}{x} +\ln x +\frac{1}{2} (1-\omega) \left(x-\ln x\right) \right. \\
    & & & \left. + \sum\limits_{k = 2}^{\omega-1} \binom{\omega-1}{k} \left(-\frac{x}{2}\right)^k \left(\frac{1}{x(1-k)}+\frac{1}{k}\right) \right]
    \end{aligned}$ & $g(T) = \dfrac{\Omega_{0} T_{0} A^{-3(1+\omega)}}{12 \rho_{c}} \dfrac{1}{x}$ \\ \hline
    
    $\mu \left(\frac{T}{T_{0}}\right)^{\sigma} \left(\frac{B}{B_{0}}\right)^{\gamma}$ & \multicolumn{2}{c|}{ No analytical solution exists} \\ \hline
    
\end{tabularx}
\caption{A summary of the reconstructed Lagrangian solutions as well as their asymptotic forms close to the bounce point for the case of a matter bounce cosmology. For simplicity, the variables $x \coloneqq 1 - \sqrt{1-\frac{T}{\rho_c}} $, $y \coloneqq \frac{12 \rho_c}{B}$ and $z \coloneqq \frac{B}{6 \rho_{c}}$ have been defined. Here, $C_{10,11,12}$ are integration constants and $n \in \mathbb{N}$, while the Green's function $G(z,s)$ as defined in  Eq.\eqref{eq:greens-func} appearing in the $Tg(B)$ model has $a(s) = 2s^2 (1-s)$ while $f(s) = 1 - \frac{T_0 \Omega_0 A^{-3(1+\omega)} s^{\omega}}{4 \rho_{c} (1-s)}$.}
\label{table: Table 4}
\end{table*}

\subsection{\label{criti_den_scale_fac} Model IV:  Matter Bounce}

The next model is one which derives from loop quantum cosmology (LQC) and generates the so called matter bounce cosmology \cite{Singh:2006im,WilsonEwing:2012pu}. This type of bouncing cosmology has been investigated during the early stages of the universe and has shown the ability to produce a scale-invariant (or nearly scale-invariant) power spectrum depending on the matter fluid considered \cite{Cai:2011tc,Cai:2009fn,WilsonEwing:2012pu,Cai:2014bea,Cai:2012va}. The scale factor which describes this type of bouncing cosmology is given by
\begin{equation}
    a(t) = A \left(\frac{3}{2} \rho_c t^2 + 1\right)^{\frac{1}{3}}\,,
\end{equation}
where $A>0$ is a constant and $0 < \rho_c \ll 1$ is a critical density which value stems from LQC. Here, $H$, $T$ and $B$ take the following forms,
\begin{align*}
    &H = \frac{2 \rho_c t}{3 \rho_c t^2 + 2}, & T = \frac{24 \rho_c^2 t^2}{(3 \rho_c t^2 + 2)^2} = \frac{2B}{3} \left(1-\frac{B}{6 \rho_{c}}\right), 
\end{align*}
\begin{equation}
B = \frac{12 \rho_c}{3 \rho_c t^2 + 2} = \frac{3T}{1-\sqrt{1-\frac{T}{\rho_c}}},
\end{equation}
and one can clearly observe that a non-singular bounce occurs at $t = 0$. The scale factor could therefore be expressed in terms of $T$ as
\begin{equation}
    a(T) = A \left[\frac{2 \rho_c}{T} \left(1 - \sqrt{1 - \frac{T}{\rho_c}}\right)\right]^{\frac{1}{3}}
\end{equation}
Defining once more a time $t_0$ where $a(t_0) = 1$ leads to
\begin{equation}
    t_0^2 = \frac{2}{3 \rho_c} \Big(\frac{1}{A^3} - 1\Big),
\end{equation}
which, since $\rho_c > 0$ imposes the condition $A < 1$. The reconstructed Lagrangians as well as their asymptotic forms close to the bounce (meaning $|T| \ll 1$ and $|B-6\rho_c| \ll 1$) are summarised in Table \ref{table: Table 4}. 

Starting off with the additive models, the $g(T)$ reconstructed solution matches with the one obtained in Ref.\cite{ElHanafy:2017sih} and with the dust case solution $(\omega = 0)$, which stems from LQC theory, as found in Ref.\cite{Bamba:2012ka}. It is also noted that the $h(B)$ asymptotic result is similar to the one obtained in unimodular $f(R)$ gravity \cite{Nojiri:2016ygo}. For the $Tg(B)$ model, only the homogeneous solutions are analytically obtained while the matter dependent solution can be only expressed in terms of an integral. However, contrary to the scenario observed in the oscillatory case, the hypergeometric functions for the given domain of $0 < z \leq 1$ are indeed complex. To avoid the presence of a complex Lagrangian, $C_{11,12}$ are set to be zero. Next, the $T+Bg(T)$ solution depends on the value of the EoS and can take on simple forms if particular values are considered. Finally, no analytic solutions have been found for the power-law ansatz.

Close to the bounce, the Lagrangian takes different asymptotic forms depending on the model. For the additive case, the Lagrangian behaves as $\Lambda$CDM with modifications while for the $Tg(B)$ model, the Lagrangian behaves as rescaled TEGR with a logarithmic correction. On the other hand, the $T+Bg(T)$ model behaves as $\mathcal{L}_\text{grav.} \propto \frac{B}{1-\sqrt{1-\frac{T}{\rho_c}}} \sim \frac{B^2}{T}$. 

Once the vacuum constraint is considered, only the $Tg(B)$ and $Bg(T)$ models satisfy the constraint as the remaining models either diverge or lead to a nonphysical zero Lagrangian. Starting with the former, the model does not require any further constraints to satisfy the vacuum constraint. In the latter $Bg(T)$ ansatz model case, however, the vacuum constraint is only possible when $\Omega_0 = 0$ and thus only generates the cosmology in the absence of other matter components if the condition is imposed. As the matter bounce cosmology is naturally constructed in the presence of dust matter, the viability of the model is questionable. 

\begin{table}[!ht]
\small
\centering
\begin{tabularx}{\columnwidth}{|x{2.5cm}|C|}
    \hline
    \multicolumn{2}{|c|}{Model IV: Matter Bounce} \\ \hline 
    
    $f(T,B)$ & Vacuum Solutions Constraints \\ \hline
    
    $g(T) + h(B)$ & $\Omega_{0} = 0$ and $C_{10} = 0$ which implies $\mathcal{L}_\text{grav.} = 0$ and therefore nonphysical \\ 
    
    $T g(B)$ & According to asymptotic behaviour, it appears to be always satisfied  \\ 
    
    $Bg(T)$ & $\Omega_0 - 0$ which implies $\mathcal{L}_{\text{grav.}} = -T + \frac{B\ln{T}}{6}$ \\
    
    $T + Bg(T)$ & $\Omega_{0} = 0$ which implies $\mathcal{L}_\text{grav.} = 0$ and therefore nonphysical\\ \hline
    
\end{tabularx}
\caption{A summary of the vacuum solution constraints for the case of matter bounce cosmology reconstruction. Overall, only the $Tg(B)$ and $Bg(T)$ model obey the constraint.}
\label{table:critical-vacuum}
\end{table}

\subsection{\label{exp_2_scale_factor}Model V: Type I--IV (Past/Future) Singularities and Little Rip Cosmologies}

The final model is an exponential scale factor of the form 
\begin{equation}\label{eq:TypeI-IV-scalefactor}
    a(t) = A \exp\left[\frac{f_0}{\alpha + 1} (t-t_s)^{\alpha + 1}\right]\,,
\end{equation}
where $A>0$ is a dimensionless constant such that $a(t_s) = A$ at the bouncing time $t_s$, $\alpha \neq -1,0,1$ is a constant\footnote{The restrictions correspond to superbounce $(\alpha = -1)$ and symmetric bounce $(\alpha = 1)$ which have been considered in previous sections. de Sitter cosmology appears for $\alpha = 0$ which case is not investigated in this work as the relevant analysis has been carried out in Refs.\cite{Bahamonde:2016cul,Zubair:2018wyy} for the case of $f(T,B)$ gravity.}, and $f_0 > 0$ is a constant with time dimension $[\text{T}]^{-(1+\alpha)}$. For times $t > t_s$, the bounce point is referred to as a past singularity while for $t < t_s$, it is referred to as a future singularity. 

For simplicity, we shift the bouncing time through a redefinition of the time coordinate $t_* = t-t_s$. Furthermore, it is assumed that the bounce represents a past singularity which now occurs at $t_* = 0$.\footnote{The analysis can be similarly repeated in the case of a future singularity. This can be achieved by setting $\alpha = \frac{2n+1}{2m+1}$ where $n, \, m \in \mathbb{Z}$, as highlighted in Refs.\cite{Nojiri:2015wsa,Odintsov:2015jca,Oikonomou:2015qha,Nojiri:2017ncd}, which ensures all cosmological parameters are well-defined.} In this way, the Hubble parameter $H$, the torsion scalar $T$ and the boundary term $B$ are given by
\begin{align}
    &H = f_0 {t_*}^{\alpha}, & &T = 6 {f_0}^2 {t_*}^{2 \alpha}, & &B = 3 T + 6 \alpha f_0 \left(\frac{T}{6 {f_0}^2}\right)^{\frac{\alpha - 1}{2 \alpha}}. \label{eq:HTB-TypeI-IV-sing}
\end{align}

Setting $t_* = t_0$ to be some time when the scale factor is unity yields the constraint
\begin{equation}
    t_{0}^{\alpha + 1} = -\frac{\alpha + 1}{f_{0}} \ln A,
\end{equation}
which imposes a constraint on the parameter $A$ depending on the magnitude of $\alpha$, namely $0<A<1$ for $\alpha > -1$ and $A >1$ for $\alpha < -1$. Ultimately, the scale factor can be solely expressed in terms of $T$ as
\begin{equation}
    a(T) = A \exp\left[\frac{f_0 t_0^{\alpha + 1}}{\alpha + 1} \left(\frac{T}{T_0}\right)^{\frac{\alpha + 1}{2 \alpha}}\right].
\end{equation}

Depending on the choice of the parameter $\alpha$, various types of bouncing cosmologies can be constructed, which are primarily classified as follows:\footnote{For a general overview of the discussed singularities, see Refs.\cite{Nojiri:2004ip,Nojiri:2004pf,Nojiri:2005sx,Cattoen:2005dx,Bamba:2009uf,Houndjo:2012ij,Odintsov:2015zza,Odintsov:2015ynk}.}
\begin{enumerate}
    \item $\alpha < -1$ (Type I/Big Rip Singularities): Characterised by a diverging scale factor and Hubble parameter at the singularity (which occurs at a finite time), these type of singularities describe an accelerated expansion which causes a dissociation of gravitationally bound structures \cite{Caldwell:2003vq}, which can be avoided through the use of dynamical fluids \cite{GonzalezDiaz:2003bc,BouhmadiLopez:2004me} or due to quantum effects \cite{Elizalde:2004mq,Nojiri:2004pf,Nojiri:2005sx}; 
    \item $\alpha > 0$ (Little Rip Cosmologies): Contrary to Type~I singularities, $a(t)$ and $H(t)$ diverge at infinite time. Nonetheless, these cosmologies still cause a dissociation of structure \cite{Frampton:2011sp}; 
    \item $0 < \alpha < 1$ (Type II/Sudden Singularities): Characterised by a diverging pressure ($\ddot{a}(t_s) \to -\infty$) \cite{Barrow:2004xh}, such universes experience a strong deceleration and had been confronted against cosmological observations \cite{Ghodsi:2011wu,Denkiewicz:2012bz}, and have been investigated in the context of closed universes \cite{Barrow:2004hk} and the resulting cosmology post the singularity \cite{FernandezJambrina:2004yy,Ghodsi:2011wu}; 
    \item $-1 < \alpha < 0$ (Type III/Big Freeze Singularity) \cite{Stefancic:2004kb}: This singularity appears when the first and higher derivatives of the scale factor diverge at the singularity \cite{Odintsov:2015zza,Borowiec:2015qrp}, and has been studied in the context for inflation due its decreasing comoving horizon close to the singularity \cite{Borowiec:2015qrp,Yurov:2007tw};
    \item $\alpha > 1$ (Type IV) \cite{Nojiri:2005sx}: Here, only the higher order derivatives of the Hubble parameter diverge i.e. $H^{(n)}(t_s) \to \infty$ for some $n \geq 2$. For such cases, the universe continues to evolve smoothly past the singularity, avoiding the need of quantum corrections \cite{Nojiri:2015wsa,Oikonomou:2015qha} and allows for a graceful exit mechanism to inflation \cite{Odintsov:2015jca}. However, it generates a variant scalar power spectrum which may be very red tilted (which could be addressed through quantum considerations) \cite{Odintsov:2015ynk,Oikonomou:2015qha}. 
\end{enumerate}

\begin{table*}[!ht]
\small
\centering
\begin{tabularx}{\textwidth}{ |c|C|C| }
    \hline
    \multicolumn{3}{|c|}{Model V: Type I--IV (Past/Future) Singularities and Little Rip Cosmologies} \\
    \hline  
    
    $f(T,B)$ & Solutions &  Asymptotic form close to the bounce \\  \hline 
    
    \multirow{2}{*}{$g(T) + h(B)$} & $g(T) = T + T_{0} \Omega_{0} A^{-3(1+\omega)} \, _1F_1 \left[-\frac{\alpha}{1 + \alpha}, \frac{1}{1 + \alpha}; - \frac{x}{1+\alpha}\right]$ & $\begin{aligned}
    &\alpha > -1 & &g(T) = T + T_0 \Omega_0 A^{-3(1+\omega)} \left(1 + \frac{\alpha}{1 + \alpha} x\right) \\
    &\alpha<-1 & &g(T) = T + \frac{T_0 \Omega_0 A^{-3(1+\omega)}}{x} \exp{\left( -\frac{x}{1+\alpha} \right)}
    \end{aligned}$ \\ 
    & \multicolumn{2}{c|}{No analytical solution for $h(B)$} \\ \hline
    
    $T g(B)$ & \multicolumn{2}{c|}{No analytical solution for $g(B)$} \\ \hline
    
    $T + Bg(T)$ & $g(T) = \frac{T_{0} \Omega_0 A^{-3(1+\omega)}}{6 T}\, _1F_1\left[-\frac{2 \alpha}{1 + \alpha}, \frac{1-\alpha}{1+\alpha}; -\frac{x}{1+\alpha} \right]$ & $\begin{aligned}
    & \alpha >-1 & &g(T) = \frac{T_0 \Omega_0 A^{-3(1+\omega)}}{6 T} \left( 1 - \frac{2\alpha}{\alpha^2 - 1} x \right) \\
    &\alpha < -1 & &g(T) = \frac{T_0 \Omega_0 A^{-3 (1+\omega)} \alpha}{3 Tx} \exp{\left(-\frac{x}{1+\alpha}\right)}
    \end{aligned}$ \\ \hline
    
    $\mu \left(\frac{T}{T_{0}}\right)^{\sigma} \left(\frac{B}{B_{0}}\right)^{\gamma}$ & \multicolumn{2}{c|}{No analytical solution exists} \\ \hline

\end{tabularx}
\caption{A summary of the Lagrangian analytical solutions as well as their corresponding asymptotic forms close to the bounce for the Hubble parameter $H(t) \propto t^\alpha$, which describes Type I--IV singularities as well as Little Rip cosmologies. Here, we have defined the parameter $x \coloneqq 3
f_0 {t_0}^{\alpha + 1} (1+\omega) \left(\frac{T}{T_0}\right)^{\frac{\alpha + 1}{2 \alpha}}$ while $_{1}F_{1}[a,b;c]$ represents the confluent hypergeometric function of the first kind which is undefined for $b \in \mathbb{Z}^- \cup \lbrace 0 \rbrace$.}
\label{table: Table 5}
\end{table*}

Based on the above considerations, the corresponding reconstructed solutions as well as the corresponding asymptotic behaviours close to the bounce are derived and summarised in Table~\ref{table: Table 5}. It is remarked that only a few analytical solutions are obtained in this case. This stems from the relationship between $T$ and $B$, Eq.\eqref{eq:HTB-TypeI-IV-sing}, which is generally not invertible\footnote{See Appendix~\ref{app:additive} for further details regarding the additive ansatz case.}. Furthermore, the solutions are not exhaustive due to the nature of the confluent hypergeometric function of the first kind. For specific choices of $\alpha$ when the function becomes undefined, one has to solve the resulting ODE on a case by case basis, whenever this is possible. 

Looking instead at the asymptotic forms of the resulting Lagrangian behaviour, the form changes according to the nature of $\alpha$. It is noted that for \mbox{$\alpha > -1$}, the Lagrangian effectively behaves as a power-law model, with $\mathcal{L}_\text{grav.} \sim \Lambda_0 + T^{\frac{\alpha+1}{2\alpha}}$, for some constant $\Lambda_0$, and $\mathcal{L}_\text{grav.} \sim \frac{B}{T} + BT^{\frac{1-\alpha}{2\alpha}}$ for the additive and $T+Bg(T)$ models respectively. In particular, for the additive $g(T)$ solution, it is observed that the asymptotic Type~III behaviour obtained in Ref.\cite{Setare:2012vs} is obtained, which also matches with the requirement that the torsion scalar exponent has to be negative \cite{Bamba:2012vg}. On the other hand, when $\alpha < -1$ (i.e. Type~I), the models effectively behave as \mbox{$\mathcal{L}_\text{grav.} \sim T^{-\frac{\alpha+1}{2\alpha}} \exp\left(T^{\frac{\alpha+1}{2\alpha}}\right)$} and \mbox{$\mathcal{L}_\text{grav.} \sim B T^{-\frac{3\alpha+1}{2\alpha}} \exp\left(T^{\frac{\alpha+1}{2\alpha}}\right)$}.

Lastly, the vacuum constraints are summarised in Table~\ref{table:TypeI-IV-sing}. It is observed that the Type~III singularity naturally satisfies the vacuum constraint. For the additive ansatz in particular, despite that the analytical homogeneous solutions for $h(B)$ are unknown, the corresponding integration constants can be set to zero allowing for the vacuum constraint to be satisfied. This, however, would reduce the model to $f(T)$ gravity. For the remaining cosmologies, only the additive model in the presence of matter may satisfy the constraint provided that $h(0) \neq 0$ as otherwise, the Lagrangian becomes identically null making the model unrealistic.  

\begin{table}[!ht]
\small
\centering
\begin{tabularx}{\columnwidth}{|x{2.3cm}|C|}
    \hline
    \multicolumn{2}{|c|}{Model V: Type I--IV Singularities} \\
    \hline 
    
    $f(T,B)$ & Vacuum Solutions Constraints\\
    \hline 
    
    $g(T) + h(B)$ & {\def\arraystretch{1.25}\tabcolsep=5pt
    \begin{tabular}{cp{4 cm}}
    $\alpha > 0, \, \alpha <-1$ & $T_0 \Omega_0 A^{-3(1+\omega)} = -h(0)$ \\ 
    $-1<\alpha<0$ & $h(0) = 0$
    \end{tabular}} \\ \hline
    
    $Bg(T)$ & {\def\arraystretch{1.25}\tabcolsep=5pt
    \begin{tabular}{cp{4 cm}}
    $\alpha > 1, \, \alpha <-1$ & $\Omega_0=0$ \\ 
    $0<\alpha <1$ & Not possible \\
    $-1<\alpha<0$ & Always satisfied
    \end{tabular}} \\ \hline
    
    $T+Bg(T)$ & $\begin{aligned}
    &\alpha > 0, \, \alpha < -1 & &\text{$\Omega_0 = 0$ but $\mathcal{L}_\text{grav.} = 0$} \\ 
    &-1<\alpha<0 & &\text{Always satisfied}
    \end{aligned}$\\ \hline
    
\end{tabularx}
\caption{A summary of the Lagrangian vacuum constraints for the different possible $\alpha$ parameter choices for the scale factor Eq.\eqref{eq:TypeI-IV-scalefactor}, which yields Type I--IV singularity and Little Rip cosmology scenarios.}
\label{table:TypeI-IV-sing}
\end{table}

\section{\label{sec:conclu}Conclusion}

In recent years, bouncing cosmologies have become attractive alternative to the inflationary paradigm, especially in the absence of initial conditions in the cosmic evolution of the Universe, as well as the possible absence of an initial singularity in the Big Bang model of cosmic expansion. Here, we have investigated the possibility of reproducing some important bouncing cosmologies within the framework of TG. In this framework, gravity is expressed as a torsional rather than curvature manifestation. As a by-product, theories constructed in this landscape are naturally lower-order meaning that the dynamical equations of GR are produced with a lower-order Lagrangian as evidenced by the appearance of a boundary term in Eq.(\ref{Ricci_torsion_equiv}).

Modified gravity is an ideal platform on which to produce new cosmological models in which longstanding cosmological problems are alleviated or entirely eliminated. One of the most popular of these models is $f(\mathring{R})$ gravity which extends GR with fourth-order contributions. In this work, we have explored the TG analog of this model, namely $f(T,B)$ gravity, which is a much broader framework to construct cosmological models due to the decoupling of the second-order torsion scalar and fourth-order boundary term. While extensions to GR \cite{Sotiriou:2008rp,Capozziello:2011et,Clifton:2011jh} have been heavily studied, their TG analog have not, and reveal interesting phenomenology beyond standard gravity.

Our approach has been to reconstruct prototype Lagrangians against well-known bouncing cosmologies in a flat FLRW background. These models may provide interesting behaviour to study the early Universe within TG. On the other hand, it is imperative that these permit Minkowski and Schwarzschild solutions to be physically viable. This is achieved by demanding that the vacuum limit (vanishing torsion scalar and boundary term) produce vacuum solutions. This vacuum condition is crucial to constructing physically admissible theories. In the following, we summarize the core results of this work and the role that this vacuum conditions plays in restricting these models.

Firstly, we considered the symmetric bouncing cosmology which is free of singularities and where the scale factor decreases to a (non-zero) minimum totally avoiding a Big Bang-like singularity. This is produced by a scale factor that decreases to this minimum and then increases after the minimum is obtained. This model produces a linear Hubble parameter when viewed as a function of cosmic time. This is shown in Fig.~\ref{fig:bounce_beha} where the energy density (pressure) approach the nonzero minimum (maximum) free of singularities. Despite being overly simplified, this represents the idea of bouncing cosmologies in a concrete way. By taking several prototype forms of the Lagrangian, the corresponding action invariant is formed in terms of the associated torsion scalar and boundary terms. As shown in Table~\ref{table: Table 1}, analytic expressions for the gravitational Lagrangian are difficult to obtain and only the additive ansatz realises the vacuum constraint.

We also explore the behavior of power-law, superbounce solutions. However, these solutions contain a singularity at their Big Bang/Crunch point in which the scale factor vanishes only to immediately re-expand, which also produces a singularity in the Hubble parameter at those points. This may be alleviated in future models by setting a minimal value for the scale factor at these times. Nevertheless, analytic solutions for this model are obtained in Table~\ref{table: Table 2} which are simpler in form and mostly obey the vacuum condition.

Similarly, this occurrence also infiltrates the oscillating bouncing solutions which can be found in Table \ref{table: Table 3}. Oscillating bouncing solutions are another representative example of bouncing cosmologies. Moreover, these solutions are more intricate in which they occasionally violate the vacuum condition. Next, we investigate the case of matter bounce cosmology which is singularity-free. This could be considered a modified power-law with more complex and realistic characteristics. Naturally, in this case the solutions are more complex as evidenced in Table~\ref{table: Table 4} which generally does not observe the vacuum conditions.

Finally, we explore the case of finite time Type I--IV singularities and Little Rip cosmologies. Here, solutions are difficult to obtain and, in the case of Type~III singularities, they mostly observe the vacuum condition. The analytical forms are shown in Table~\ref{table: Table 5}.

This exploration of bouncing solutions within the $f(T,B)$ gravity framework may open the door to future work on early Universe cosmology stemming from this scenario of gravity. In this work, we have investigated the potential model that may emerge for bouncing solutions at the level of background cosmology. To further restrict physically relevant models, we need to study the early Universe perturbations of each of these models and to investigate their impact on the cosmic microwave background. This would be interesting and may illuminate particular features of TG.

\appendix

\section{\texorpdfstring{$f(T,B) = g(T) + h(B)$}{f(T,B) = g(T) + h(B)}}
\label{app:additive}

In the case of separable additive models, similar to other works in reconstruction Refs.\cite{delaCruzDombriz:2011wn,Bahamonde:2016cul,Zubair:2018wyy}, the Friedmann equation Eq.\eqref{eq:friedmann_mod} yields a separable partial differential equation which can yield solutions provided that there exists in invertable relation between the coordinate time $t$, the torsion scalar $T$ and the boundary term $B$. This means that either the matter sector can be described either through $T$ or $B$, or that Eq.\eqref{eq:friedmann_mod} can be solely expressed in terms of either variable. The latter case is not considered since this would result in a complicated expression which is difficult to solve analytically. Furthermore, this means that we would be investigating the case where $f(T,B) = f(T) = f(B)$ which is not of interest here. Therefore, we only investigate the former case. Without loss of generality, it shall be assumed that the matter sector is sourced by the $g(T)$ function. Notwithstanding, if modifications to TEGR were to be considered, meaning $g(T) = 0$, $h(B)$ would act as the source for describing the resulting cosmological behaviour. Such considerations have been considered in Refs.\cite{Bahamonde:2016cul,Zubair:2018wyy,Paliathanasis:2017flf} and thus are not explored in further detail here. 

For this ansatz, Eq.\eqref{eq:friedmann_mod} gives the following system of separable equations:
\begin{align}
    &g - 2T g_T + T = T_0 \Omega_0 a(T)^{-3(1+\omega)}, \\
    &h-B h_B + 6H \dot{B} h_{BB} = 0.
\end{align}
The solvability of the above system ultimately depends on whether the scale factor can be expressed in terms of the torsion scalar and whether the coefficient of $h_{BB}$ can be solely expressed in terms of $B$.

As an illustrative working example, in the case of an oscillatory scale factor, the system reduces to
\begin{align}
    &g - 2 T g_T + T = T_0 \Omega_0 A^{-3(1+\omega)} \left(1 + \frac{t_* T}{24 C^2}\right)^{3(1+\omega)}, \\
    &h - B h_B - \frac{2}{5} (B + y) (B + 6 y) h_{BB} = 0,
\end{align}
which yields the solutions as summarised in Table~\ref{table: Table 3}.

However, in the case of Type I--IV singularities, the $h(B)$ solution cannot be obtained since the $h_{BB}$ coefficient, which is given to be
\begin{equation}
    \frac{B- 3T}{\alpha} \left[(\alpha - 1) (B - 3T) + 6T\right],
\end{equation}
cannot be expressed solely in terms of $B$. This stems from the relation between the $B$ and $T$ scalars Eq.\eqref{eq:HTB-TypeI-IV-sing}
\begin{equation}
    B = 3T + 6\alpha f_0 \left(\frac{T}{6{f_0}^2}\right)^{\frac{\alpha-1}{2\alpha}},
\end{equation}
which cannot be inverted in general. In principle, one may able to solve $h(B)$ for certain special cases where the relation becomes invertable, for example $\alpha = -\frac{1}{3}$. However, no analytical solution was found for this particular case.

\section{\texorpdfstring{$f(T,B) = T g(B)$}{f(T,B) = T g(B)}}
\label{app:Greens-Method}

For these ansatz models, for most of the bouncing models which have been investigated, Frobenius method was applied to obtain the homogeneous solutions of the resulting differential equation, which is generally expressed in the form
\begin{equation}
    \alpha(z)g''(z) + \beta(z) g'(z)+\gamma(z) g(z) = f(z),
\end{equation}
for some functions $\alpha, \, \beta, \, \gamma,$ and $f(z)$. Namely, this method assumes a solution of the form
\begin{equation}
    g_\text{hom.}(z) = \sum_{n=0}^{\infty} D_n z^{n+r},
\end{equation}
where $D_n$ are coefficients and $r$ is a constant which is determined from the resulting differential equation. Once the homogeneous solutions are obtained, the particular solution could then be derived through the use of the Wronskian and Green's function as follows \cite{kythe2011green}: if $g_{1,2}$ represent the homogeneous solutions, the particular solution is obtained from
\begin{equation}
    g_\text{part.}(z) = \int G(z,s) f(s) \, ds, 
\end{equation}
where $G(z,s)$ is the Green's function defined to be
\begin{equation}\label{eq:greens-func}
    G(z,s) = \frac{g_2(z)g_1(s)-g_1(z)g_2(s)}{\alpha(s)W(s)}
\end{equation}
with $W(s) = g'_2 g_1 - g'_1 g_2$ being the Wronksian.

As an example, we illustrate the procedure for the symmetric bouncing model. In this case, the Friedmann equation reduces to
\begin{align}
    g(z)& + (z - y) g'(z) - 2 y z g''(z) \\ \nonumber
    &= 1 - \frac{3T_{0} \Omega_{0} A^{-3(1+\omega)}}{z} \exp\left(- \frac{(1+\omega) z}{y}\right),
\end{align}
where $z = B-y$ and $y = \frac{12 \alpha}{t_*^2}$. Solving using Frobenius' approach yields two independent solutions for $r=0$ and $r=\frac{1}{2}$, which are
\begin{align}
    g_1(z) &= D_{1} \sum_{n=0}^{\infty} \frac{z^{n}}{(2n-1)!!\, y^n} \nonumber \\
    &= D_{1} \left[1+\sqrt{\frac{z\pi }{2y}} e^{\frac{z}{2y}} \text{erf}\left(\sqrt{\frac{z}{2y}}\right)\right], \\
    g_2(z) &= D_{2} \sqrt{z} \sum_{n=0}^{\infty} \frac{z^{n}}{(2n)!!\, y^n} = D_{2} \sqrt{z} e^{\frac{z}{2y}}.
\end{align}
where $D_{1,2}$ are constants determined by boundary conditions. The particular solution turns out to be 
\begin{equation}
    g_\text{part.}(z) = 1 + \int\limits^z G(z,s) h(s) \, ds,
\end{equation}
with \mbox{$h(s) = - \frac{3 T_{0} \Omega_{0} A^{-3(1+\omega)}}{s} \exp\left(- \frac{(1+\omega) s}{2 y}\right)$} and \mbox{$\alpha(s) = -2sz$}, for which, the general integral solution is not obtained for arbitrary values of $\omega$. Nonetheless, the integral can be evaluated close to the bounce where $|z| \ll 1$, which yields
{\small
\begin{align}
    g_\text{part.}(z) &= 1 -\frac{2 R}{y} -\frac{R (y+z) \ln z}{y^2} -\frac{R (1+\omega) z}{2 y^2} + \mathcal{O}(z^2) \nonumber \\
    &\approx 1 -\frac{R \ln z}{y}
\end{align}
}
where $R \coloneqq 3T_0 \Omega_0 A^{-3(1+\omega)}$.

\section{\texorpdfstring{$f(T,B) = \mu \left(\frac{T}{T_0}\right)^{\sigma}\left(\frac{B}{B_0}\right)^{\gamma}$}{f(T,B) = mu (T/T0)sigma (B/B0)gamma}}
\label{app:power-law}

For these models, as the form of the Lagrangian is already assumed, only the parameters $\sigma$ and $\gamma$ are to be constrained. In general, the Friedmann equation to solve is
\begin{align}
    &T = T_0 \Omega_0 a(T)^{-3(1+\omega)} + \mu \left(\frac{T}{T_0}\right)^{\sigma}\left(\frac{B}{B_0}\right)^{\gamma} \left[2\sigma + \gamma -1 \right.\nonumber \\
    &\left.-2\gamma \sigma + \frac{6\gamma \sigma T}{B} - 6H\gamma(\gamma-1)\frac{\dot{B}}{B^2}\right].
\end{align}
Once a bouncing model is chosen, the Friedmann equation could then be solely expressed in terms of $T$ and $B$ which could then be used to determine the parameters. Here, we illustrate the procedure for the symmetric bouncing cosmology as an example, for which the Friedmann equation takes the form
\begin{align}
    &T = T_0 \Omega_0 a(T)^{-3(1+\omega)} + \mu \left(\frac{T}{T_0}\right)^{\sigma}\left(\frac{B}{B_0}\right)^{\gamma} \left[2\sigma + \gamma -1 \right.\nonumber \\
    &\left.-2\gamma \sigma + \frac{6 \gamma (\sigma - \gamma + 1) T}{B} + \frac{18 \gamma (\gamma - 1) T^{2}}{B^{2}}\right].
\end{align}
Observe that when $t = 0$, $a(T) = A$, $H = T = 0$ and $B = \frac{12\alpha}{{t_*}^2}$. This means that the Lagrangian is non-singular provided $\sigma >0$. With this in mind, one arrives at the conclusion that 
\begin{equation}
    \Omega_0 A^{-3(1+\omega)} = 0.
\end{equation}
Since $A, \, \Omega_0 \neq 0$, this constraint imposes the condition that this ansatz model can only describe the symmetric bouncing cosmology when the universe is devoid of matter content and is only described by the effective torsional fluid.

Next, the value for $\mu$ can be easily computed by evaluating the resulting Friedmann expression at $t = t_0$, one can obtain an expression for the constant $\mu$ as follows
\begin{align}
    \mu &=\frac{T_0}{2\sigma + \gamma -1 - 2\gamma\sigma + \frac{6 \gamma (\sigma - \gamma + 1) T_{0}}{B_{0}} + \frac{18 \gamma (\gamma - 1) T_{0}^{2}}{B_{0}^{2}}} \equiv \frac{T_0}{\nu}
\end{align}
provided that $\nu \neq 0$. Indeed, if this were the case, this would imply that $T_0 = 0$ which contradictions the notion of the time $t = t_0$. 

By applying all the necessary conditions and the definition of $\mu$, the Friedmann equation simplifies to
\begin{align}
    &\nu = \left(\frac{T}{T_0}\right)^{\sigma-1}\left(\frac{B}{B_0}\right)^{\gamma} \left[2\sigma + \gamma -1 -2\gamma \sigma \right.\nonumber \\
    &\left. + \frac{6 \gamma (\sigma - \gamma + 1) T}{B} + \frac{18 \gamma (\gamma - 1) T^{2}}{B^{2}}\right].
\end{align}
Finally, to determine the values of $\sigma$ and $\gamma$, we require the equation to hold at any time. However, no parameter choice is able to satisfy the Friedmann equation meaning that this model cannot host symmetric bouncing cosmologies.

\end{document}